\documentclass[prb,superscript address,twocolumn]{revtex4}

\newcommand{\mbo}{$\mu_\mathrm{B}$}

\newcommand{\dos}{density of states}
\def\bra#1{\mathinner{\langle{#1}|}}
\def\ket#1{\mathinner{|{#1}\rangle}}

\usepackage[T1]{fontenc}
\usepackage{widetable}
\usepackage{color}
\usepackage{amsmath}    
\usepackage[pdftex]{graphicx} 
\DeclareGraphicsExtensions{.pdf}
\begin{document}

\title{Iron porphyrin molecules on Cu(001): Influence of adlayers and ligands on the magnetic properties}
\author{H. C. Herper}
\email[Corresponding author. Electronic address:]{heike.herper@uni-due.de}
\affiliation{Fakult\"at f\"ur Physik und Center for Nanointegration Duisburg-Essen (CENIDE), Universit\"at Duisburg-Essen, Lotharstra\ss{}e 1, 47048 Duisburg, Germany}
\affiliation{Department of Physics and Astronomy, Uppsala University, Box 516, 75120 Uppsala, Sweden}
\author{M. Bernien}
\affiliation{Institut f\"ur Experimentalphysik, Freie Universit\"at Berlin, Arnimallee 14, 14195 Berlin, Germany}
\author{S. Bhandary}
\affiliation{Department of Physics and Astronomy, Uppsala University, Box 516, 75120 Uppsala, Sweden}
\author{C. F. Hermanns}
\affiliation{Institut f\"ur Experimentalphysik, Freie Universit\"at Berlin, Arnimallee 14, 14195 Berlin, Germany}
\author{A. Kr\"uger} 
\affiliation{Institut f\"ur Experimentalphysik, Freie Universit\"at Berlin, Arnimallee 14, 14195 Berlin, Germany}
\author{J. Miguel} 
\affiliation{Advanced Light Source, Lawrence Berkeley National Laboratory, Berkeley, CA 94720, USA}
\author{C. Weis}
\affiliation{Fakult\"at f\"ur Physik und Center for Nanointegration Duisburg-Essen (CENIDE), Universit\"at Duisburg-Essen, Lotharstra\ss{}e 1, 47048 Duisburg, Germany}
\author{C. Antoniak}
\affiliation{Fakult\"at f\"ur Physik und Center for Nanointegration Duisburg-Essen (CENIDE), Universit\"at Duisburg-Essen, Lotharstra\ss{}e 1, 47048 Duisburg, Germany}
\author{B. Krumme}
\affiliation{Fakult\"at f\"ur Physik und Center for Nanointegration Duisburg-Essen (CENIDE), Universit\"at Duisburg-Essen, Lotharstra\ss{}e 1, 47048 Duisburg, Germany}
\author{D. Bovenschen}
\affiliation{Fakult\"at f\"ur Physik und Center for Nanointegration Duisburg-Essen (CENIDE), Universit\"at Duisburg-Essen, Lotharstra\ss{}e 1, 47048 Duisburg, Germany}
\author{C. Tieg}
\affiliation{European Synchrotron Radiation Facility, PB 220, 38043 Grenoble, France}
\author{ B. Sanyal}
\affiliation{Department of Physics and Astronomy, Uppsala University, Box 516, 75120 Uppsala, Sweden}
\author{E. Weschke}
\affiliation{Helmholtz-Zentrum Berlin, Institut f\"ur komplexe magnetische Materialien, 14109 Berlin, Germany}
\author{C. Czekelius}
\affiliation{Institut f\"ur Chemie und Biochemie, Freie Universit\"at Berlin, 14195 Berlin, Germany}
\author{W. Kuch}
\affiliation{Institut f\"ur Experimentalphysik, Freie Universit\"at Berlin, Arnimallee 14, 14195 Berlin, Germany}
\author{H. Wende}
\affiliation{Fakult\"at f\"ur Physik und Center for Nanointegration Duisburg-Essen (CENIDE), Universit\"at Duisburg-Essen, Lotharstra\ss{}e 1, 47048 Duisburg, Germany}
\author{O. Eriksson}
\affiliation{Department of Physics and Astronomy, Uppsala University, Box 516, 75120 Uppsala, Sweden}

\pacs{71.15.Mb, 75.30.Gw, 78.70.DM, 33.15.Kr}

\begin{abstract}The structural and magnetic properties of Fe octaethylporphyrin (OEP) molecules on Cu(001)  have been investigated by means of density functional theory (DFT) methods and X-ray absorption spectroscopy. The molecules have been adsorbed on the bare metal surface and on an oxygen-covered surface, which shows a $\sqrt{2}\times2\sqrt{2}R45^{\circ}$ reconstruction. In order to allow for a direct comparison between magnetic moments obtained from sum-rule analysis and DFT we calculate the dipolar term $7\langle T_z\rangle$, which is also important in view of the magnetic anisotropy of the molecule. The measured X-ray magnetic circular dichroism shows a strong dependence on the photon incidence angle, which we could relate to a huge value of  $7\langle T_z\rangle$, e.g. on Cu(001)  $7\langle T_z\rangle$ amounts to -2.07\,\mbo{} for normal incidence leading to a reduction of the effective spin moment $m_s + 7\langle T_z\rangle$.
Calculations have also been performed to study the influence of possible ligands such as Cl and O atoms on the magnetic properties of the molecule and the interaction between molecule and surface, because the experimental spectra display a clear dependence on the ligand, which is used to stabilize the molecule in the gas phase. Both types of ligands weaken the hybridization between surface and porphyrin molecule and change the magnetic spin state of the molecule, but the changes in the X-ray absorption are clearly related to residual Cl ligands. \end{abstract}

\maketitle
\section{Introduction} \label{sec:intro}
\begin{figure}[t]
\includegraphics[width=0.27\textwidth]{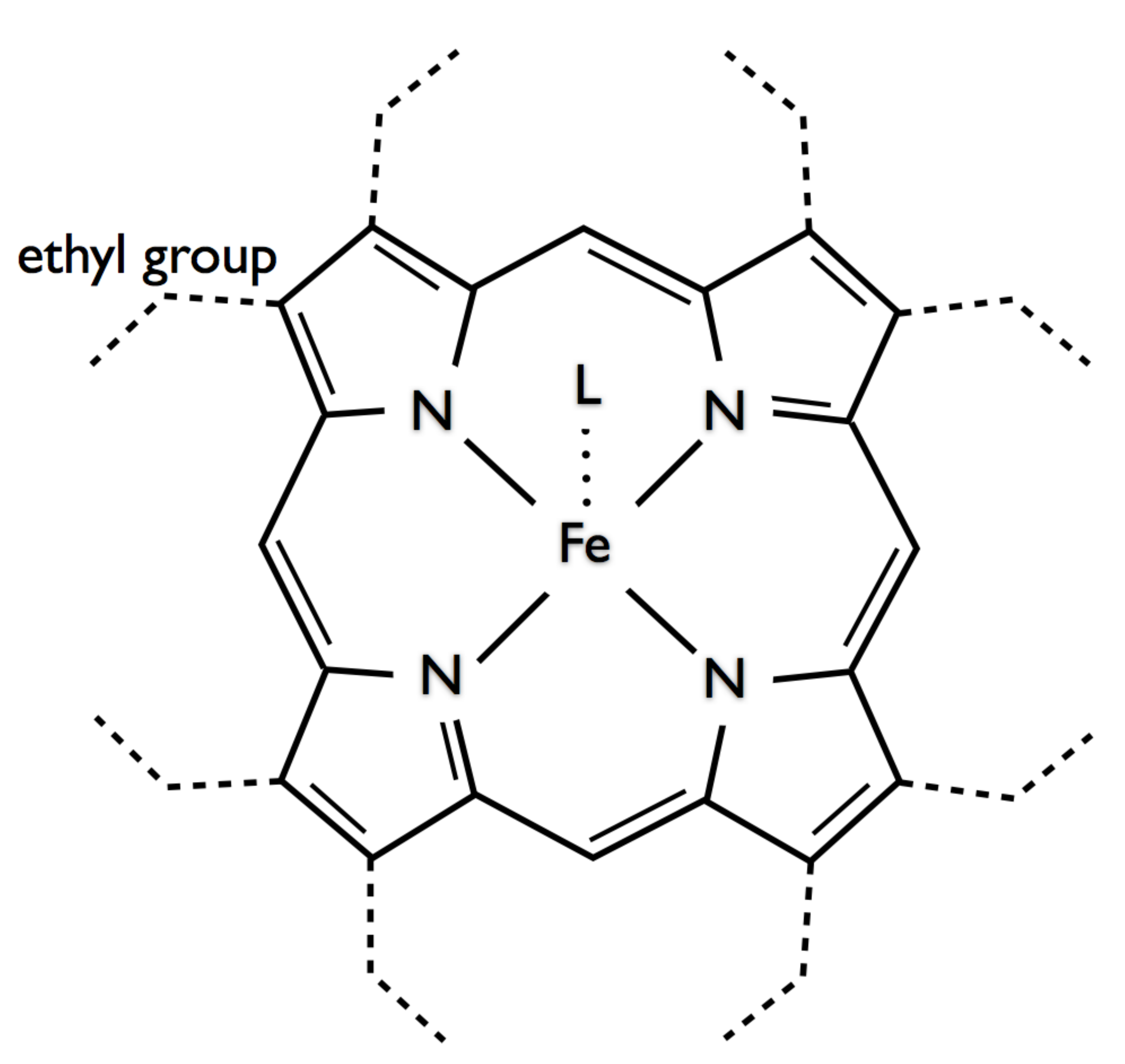}\caption{Schematic structure of the Fe OEP molecule. Hydrogen atoms are not shown. For the calculations the ethyl groups CH$_3$ are replaced by H atoms. Ligands are denoted by L  and  stick out of the molecular plane.} 
\label{fig:struc}
\end{figure}
\begin{figure*}[t]
\includegraphics[width=0.85\textwidth]{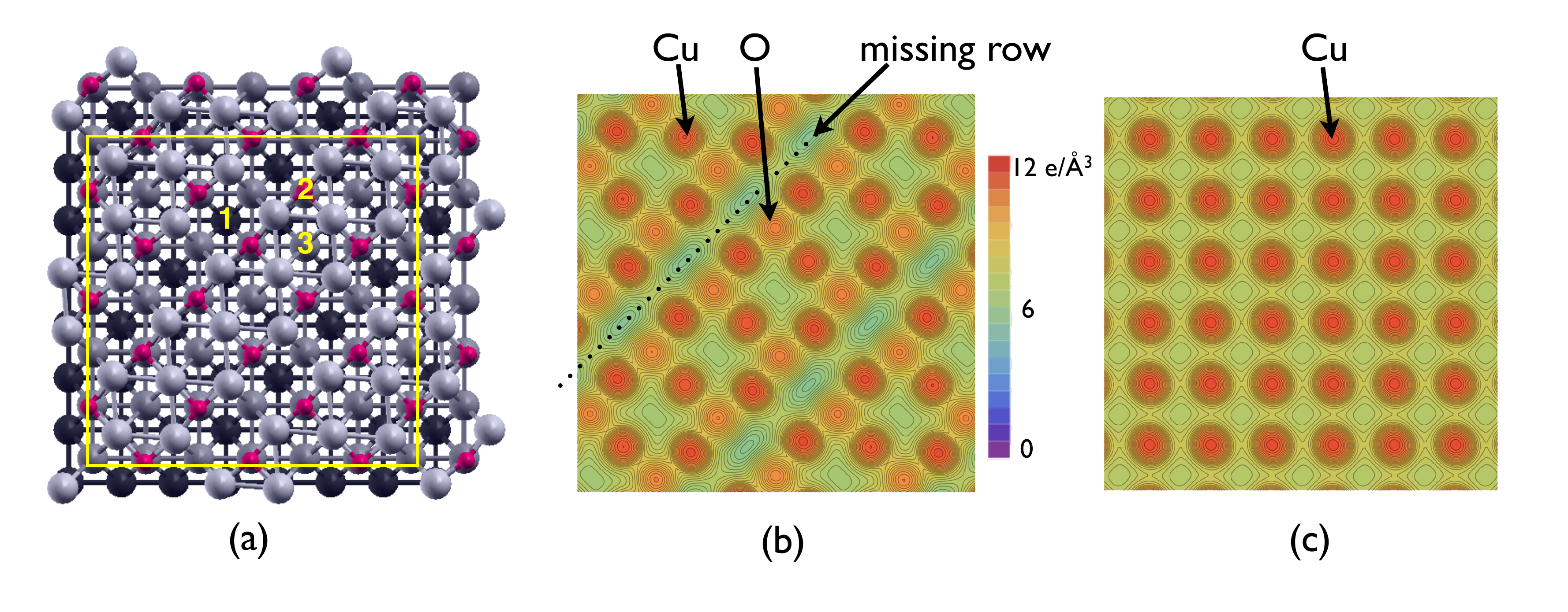}\caption{(Color online) (a) $\sqrt{2}\times2\sqrt{2}{\rm R}45^{\circ}$ missing-row reconstruction of the  Cu(001) surface with 0.5\,ML O. Cu atoms are shown in different gray shades (light gray corresponds to the surface layer and black to the fixed bottom layer.) The smaller (red) circles indicate the O atoms.  Numbers mark the different positions of the Fe center of the FeP, namely missing-row (1), on top of O (2), and hollow site (3) position. (b) Charge density plot of the reconstructed O/Cu(001) surface in comparison with the plain Cu(001) surface (c). Dark (red) circles  correspond to the charge distribution of the Cu atoms, lighter (orange) spots in (b) indicate the charge of the surface O-atoms. The bright (greenish) areas in (b) (marked by the dotted line) show the charge depletion due to the missing-row reconstruction. } 
\label{fig:OCu}
\end{figure*}

Hybrid systems of paramagnetic molecules and metallic surfaces are new promising materials for magneto-electronic devices and may in future replace devices with artificial nano-structures.\cite{Dediu:09,Leoni:11}
The ongoing search for new materials which can be used in magnetic switching devices has brought phthalocyanine and porphyrin molecules with transition metal center into focus, because they arrange themselves flat on surfaces and are relatively easy to handle experimentally under UHV conditions. Recent experiments 
with Co(II) tetraphenylporphyrin (CoTPP) molecules on Ni/Cu(001) have shown that the spin state of the Co atom can be controlled by NO ligands, which are attached or removed by temperature treatment.\cite{Waeckerlin:10} Miguel {\em et al.} tailored  the coupling  between  Fe(II) octaethylporphyrin (Fe OEP) molecules and an oxygen covered Ni/Cu(001) substrate  by the same type of ligands.\cite{Miguel:11}
It has been observed that in the presence of an intermediate oxygen layer the coupling between the paramagnetic molecules and the ferromagnetic (FM) surface is of antiferromagnetic nature, whereas the investigated systems couple ferromagnetically without oxygen.\cite{Bernien:09, Wende:07} Theoretical investigations have revealed that  Fe OEP molecules are chemisorbed at ferromagnetic surfaces whereby the magnetic coupling is mediated by the N-atoms.\cite{Wende:07} 

Transition metal-based porphyrin molecules have been studied intensively on ferromagnetic substrates focussing on the magnetic coupling between surface and molecule\,\cite{Bernien:09,Wende:07, Oppeneer:09,Chylarecka:11,Santos:12,Annese:11} and the influence of the anisotropy of the substrate on the magnetic properties of the molecule. Nonmagnetic substrates have been used to study spin and orbital magnetic anisotropies\,\cite{Stepanow:10,Stepanow2:10} as well as correlation and hybridization effects with the substrate.\cite{Stepanow:11,Fanetti:11}
 Here, we  use a non-magnetic substrate to unravel the magnetic anisotropy of the molecule and it's dependence on ligands and surface oxidation, whereby special focus is on the importance of the dipole term $7\langle T_z\rangle$.  We address the issue of how the magnetic properties like magnetic anisotropy, and the hybridization between molecule and substrate change if the FM film is absent, i.e., we study the adsorption of Fe porphyrin molecule on a bare Cu(001) substrate and pursue the question whether an O-interlayer has an influence on the magnetic properties as in the presence of an FM film. This has been done  by performing X-ray absorption measurements and density functional theory (DFT) calculations using the VASP code.\cite{Kresse:96} For a proper description of molecular bonding  van der Waals forces are important. However, they are not included in the standard DFT description, therefore,  we have used the semi-empirical  form of Grimme\,\cite{Grimme:06} to account for van der Waals interaction.  The use of a fully {\em ab initio} van der Waals term is not feasible for large systems as in the present case. 

Parallel to the present {\em ab initio} investigations  X-ray absorption experiments  have been carried out for Fe OEP on plain Cu(001) and in presence of an oxygen interlayer, which goes along with a $\sqrt{2}\times2\sqrt{2}$R45$^{\circ}$ reconstruction of the Cu surface, see Fig.\,\ref{fig:OCu}(a). We have performed angle-dependent X-ray absorption spectroscopy (XAS) and X-ray magnetic circular dichroism (XMCD) experiments in high magnetic fields at low temperatures. From these measurements we can evaluate the effective spin moments via sum-rules derived by  Carra {et al.}\cite{Carra:93} In order to compare the effective spin moments extracted from experiment and the calculated spin moments we determine the dipolar term which connects the two quantities according to the work of van der Laan.\cite{vanderLaan:98}

In the present DFT calculations the Fe OEP molecules are  modeled without the outer ethyl groups and are passivated with H atoms only, referenced as FeP in the following, see Fig.\,\ref{fig:struc}. 
Furthermore,  in the  experiment Fe OEP molecules are usually decorated with additional ligands to stabilize them. In our case we use two different types of ligands: pyridine (C$_5$H$_5$N named Py in the following), which decouples from the molecule during degassing of the molecular powder in vacuum prior to the sample preparation which yields Fe(II)OEP in its pure form and atomic Cl for which scanning tunneling microscopy measurements indicate that the Cl may partially remain in the system. We will show that the choice of the ligand, which stabilizes the molecule in air has a drastic influence on the magnetic properties of the molecule.

The paper is organized as follows. After a brief presentation of the theoretical methods including a discussion of the preparation of the surfaces in Sec.\,\ref{sec:theo}, in Sec.\,\ref{sec:exp-details} the experimental set up and sample preparation will be explained. A discussion of the results can be found in Secs.\,\ref{sec:exp} and \ref{sec:fep} where magnetic and structural properties of the FeP on the two different surface structures $-$ plain Cu(001) and oxidized Cu(001) surface $-$ as well as the influence of Cl and O ligands will be considered. Finally, a conclusion and an outlook are given in the last section (Sec.\,\ref{sec:concl}).
\section {Computational methods and surface reconstruction} \label{sec:theo}
We have performed  density functional theory (DFT) calculations employing the VASP code\,\cite{Kresse:96} with the projected augmented wave potential (PAW) method.\cite{Bloechl:94} The exchange correlation functional has been described via the generalized gradient approximation (PW91)\,\cite{PW:91,PW:91-E} plus a Hubbard-$U$ correction in order to incorporate the localized character of the $d$ orbitals of the Fe-atom in the molecule and to  get a reasonable HOMO-LUMO gap. Here, we make use of the method of Dudarev\,\cite{Dudarev:98}, where the Coulomb ($U$) and exchange ($J$) interaction enter in the Hamiltonian as an effective value, whereby  $U-J= 3.0$\,eV has been used for the Fe $d$ orbitals. The value has been previously shown to describe FeP properly.\cite{Panchmatia:08,Panchmatia:13} However, the description of molecular systems in DFT does not only suffer from  the failure of the description of correlation effects, but also does not account for van der Waals forces. Recently, several attempts have been made to include van der Waals interaction in DFT.\cite{Tkatchenko:09,Atodiresei:09,Grimme:06} For such large systems as in the present study we use the semi-empirical method of Grimme\,\cite{Grimme:06} as implemented in VASP. Fully {\em ab initio} methods are to expensive for the systems under consideration. However, a semi-empirical approach should already significantly improve the description of the molecular bonding.
All calculations have been performed for the $\Gamma$-point and a cut-off energy of 400\,eV for the plane-waves.
In order to study the magnetic and electronic properties of FeP on Cu(001) with and without an intermediate oxygen layer we have used an $8\times8$ lateral unit cell and 3 monolayer (ML) of Cu(001). The bottom layer is kept fixed to simulate bulk-like behavior. The rest of the system is fully relaxed and the vacuum size is about 14\,\AA.

If Cu(001) is covered by 0.5\,ML of oxygen this leads to a $\sqrt{2}\times2\sqrt{2}\,{\rm R}45^{\circ}$ surface reconstruction, see Fig.\,\ref{fig:OCu} (a).\cite{Zeng:89, Robinson:90}  After relaxation the O atoms are embedded almost in the surface layer being only 0.16\,\AA ~above the Cu atoms. This is in good agreement with results from low energy electron diffraction (LEED) measurements by Zeng {\em et al.}\cite{Zeng:89} This reconstruction has significant influence on the electronic structure of the surface layer.
Figure\,\ref{fig:OCu}   shows the charge distribution of the  Cu(001) $\sqrt{2}\times2\sqrt{2}\,{\rm R}45^{\circ}$ surface (b) in comparison to the plain Cu surface (c). The charge depletion along the missing-rows (blue-green areas) is clearly visible. Due to the missing-row reconstruction the electronic structure of the  O-terminated surface is more complex compared to the plain Cu(001) surface and offers a number of additional adsorption positions for the molecule, e.g., the Fe sitting on the missing-row or on top  of an oxygen atom, {\em cf.} Fig.\,\ref{fig:OCu}(a).

The surface or the ligands may not only influence the magnetic spin state but also the magneto-crystalline anisotropy (MCA), i.e., the energy needed to switch the magnetic orientation from the easy axis to the hard axis. Here, we will consider only the spin-orbit coupling induced magnetic anisotropy, for details see Ref.\,\onlinecite{Stoehr:06}. The perturbation of the original system-originating  spin-orbit coupling can be written as
  \begin{eqnarray}
    H_{SO}\,=\,\zeta(r) {\bf L\cdot S}\,=\, \zeta(r)(L_{x}S_{x}+L_{y}S_{y}+L_{z}S_{z}) \\
             \zeta=\frac{1}{4c^2r}\frac{\delta V}{\delta r}
  \end{eqnarray}
  with $\zeta$ being the spin-orbit coupling strength. For 3$d$ transition metals, $\zeta$ is in the range of 50-70 meV which is small compared to the usual band width and hence, the approximation is reasonable. Since the first order term vanishes, the first significant contribution comes from the 2nd order corrections. For 3$d$ metals, higher order terms are ignored.
Therefore, the spin-orbit contribution to the energy, from the 2nd order perturbation theory reads 
   \begin{eqnarray}\label{eq:SO}
              E_{SO}\,=\,-\zeta^2\sum_{u,o}n_{u}n_{o}\frac{[ \bra{u}{\bf L\cdot S}\ket{o} \bra{o}{\bf L\cdot S}\ket{u}]}{E_{u}-E_{o}}.
  \end{eqnarray}
Here $\ket{o}$ and $\ket{u}$ correspond to occupied and unoccupied states, respectively, in the basis of $\ket{lm,\sigma}$, where $l,m,\sigma$ are orbital, magnetic, and spin quantum numbers, respectively. $E_u$, $n_u$ and $E_o$, $n_o$ are energy eigenvalues and occupations of unoccupied and occupied levels. 
The electronic states of our interest are antibonding states of Fe-N $p-d$ hybridization arising mostly from Fe $d$ states in FeP.  The contribution to the MCA is inversely proportional to $\Delta E = E_u-E_o$ (see Eq.\,\ref{eq:SO}). Thus states close to the Fermi energy ($E_{\rm F}$) are the most important ones. The eigenvalues are taken from our calculated orbital-projected density of states and the 2nd order perturbation has been calculated for  magnetization axis along (001) and (100), i.e out-of-plane and in-plane directions.
\begin{figure}[b]
\includegraphics[width=0.3\textwidth]{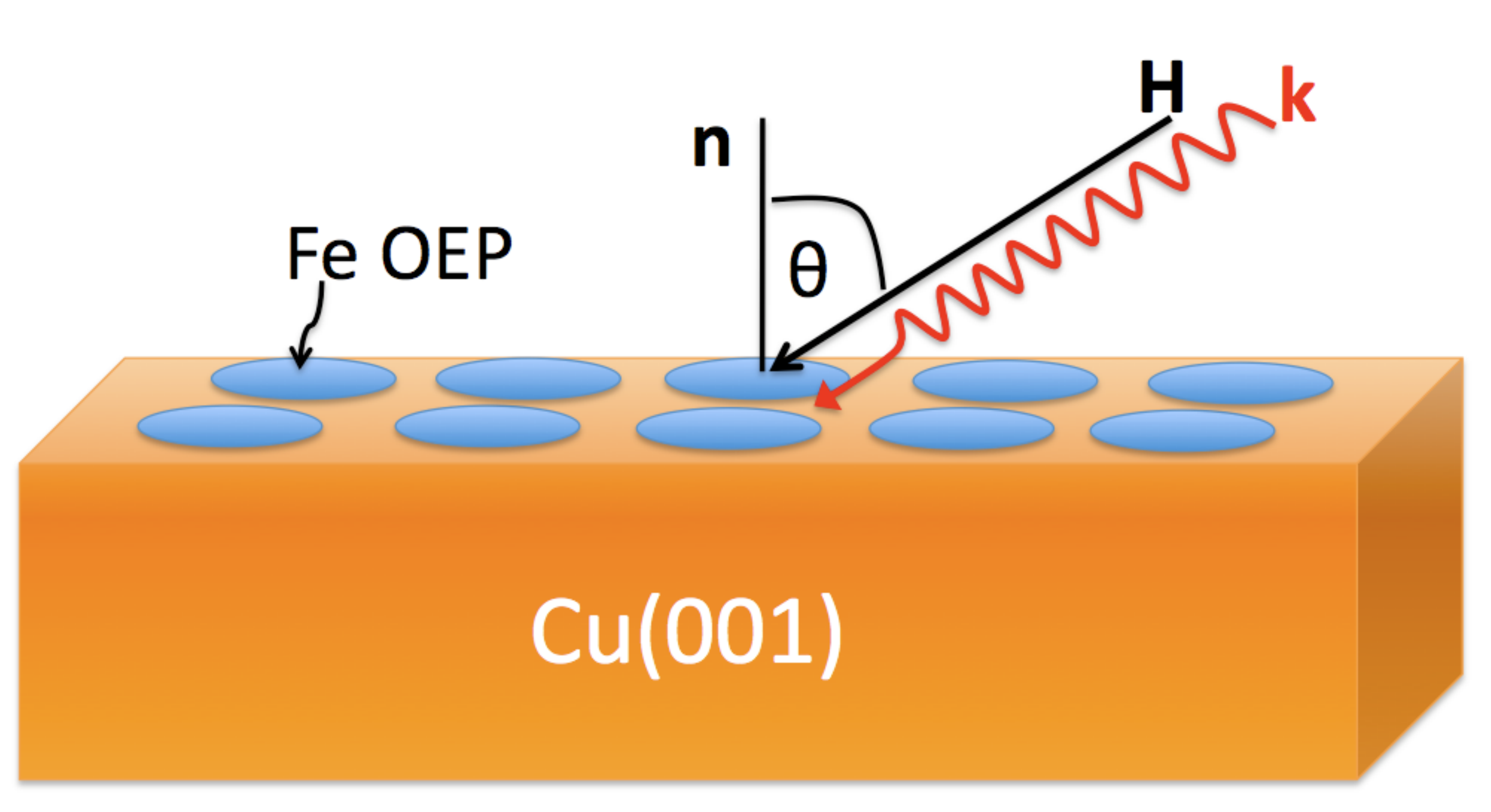}
\caption{(Color online) Sketch of the experimental setup. The molecules (ovals) are placed on the  Cu(001) surface.  The X-ray absorption measurements are carried out for different angles $\theta$ between the surface normal and the photon beam ($\bf k$). The magnetic field is always oriented parallel to the $\bf k$-vector of the photon beam and  the electric field, which probes the $d$ orbitals, is in the plane of incidence, perpendicular to the photon beam (p-polarized).} 
\label{fig:beam}
\end{figure}
\section{Experimental details}\label{sec:exp-details}
\begin{figure}[t]
\includegraphics[width=0.38\textwidth]{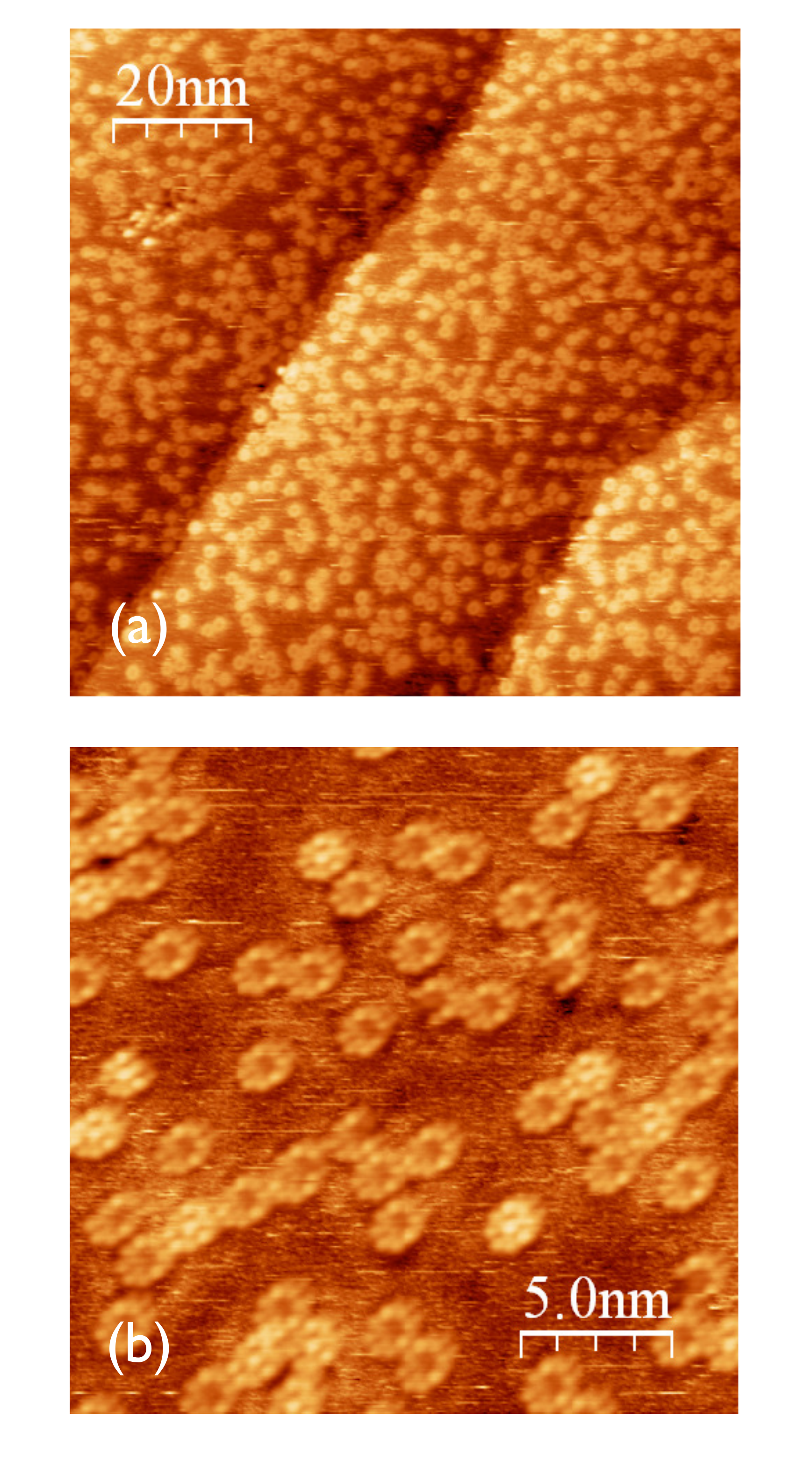}
\caption{(Color online) STM images of 0.4\,ML of Fe OEP(Cl) on Cu(100) at room temperature (a): 1.3\,V, 0.08\,nA; (b): 1.2\,V, 0.17\,nA.}
\label{fig:STM}
\end{figure}

Iron OEP  was synthesized by reduction of Fe OEP(Cl) using hydrazine and isolated as its dipyridine complex Fe OEP(Py)$_2$.\cite{Dolphin:76} Differential Scanning Calorimetry (DSC) studies of the isolated material showed loss of the pyridine ligands at 146-154\,$^{\circ}$C.\cite{note-Sams}  No further transitions were detected up to 350\,$^{\circ}$C.
Paramagnetic Fe OEP was deposited in UHV on Cu(001) and on an oxygen-reconstructed Cu(001) surface (held at room temperature) from a tantalum crucible at about 200\,$^{\circ}$C. Prior to the deposition, Fe OEP(Py)$_2$ was degased in UHV at 200\,$^{\circ}$C for about 5\,h to ensure the loss of the pyridine ligands. After deposition, N $K$ edge  XA spectra displayed no $\pi^*$ resonance in the energy range from 397\,eV to 403\,eV in addition to the resonances of the porphyrin macrocycle, showing that pyridine is not present on the surface. For the 2nd set of samples Fe OEP(Cl) was purchased from Porphyrin Systems and deposited from a tantalum crucible at about 200\,$^{\circ}$C without further treatment. Angle dependent XAS  and XMCD measurements have been performed at the beam lines UE46 PGM1 at BESSY II (samples with Py) and ID08 at the ESRF (samples with Cl) in an applied magnetic field (temperature) of  5.9\,T (5\,K) at BESSY II and 5.0\,T (8\,K) ESRF, respectively. The XAS signal was detected by means of total electron yield and normalized to a reference signal upstream to the experiment. The $\bf k$ vector of the X rays was aligned parallel to the magnetic field for different incidence angles $\theta$, see Fig.\,\ref{fig:beam}.  The third (first) harmonic of the undulator with an energy resolution set to 150~meV (250~meV) and a circular polarization degree of 85\% (100\%) was used for the measurements performed at BESSY II (ESRF). Coverages of 0.4\,ML were chosen to ensure that only molecules that are in direct contact with the substrates contribute to the XAS signal. In Fig.\,\ref{fig:STM} scanning tunneling microscopy (STM) images of 0.4\,ML of Fe OEP(Cl) on Cu(001) taken directly before the XAS measurements at ID08 are shown. It is obvious that the porphyrin molecules adsorb flat on the surface. Bright and dark centers of the molecules may be attributed to Cl ligands on top of the iron centers and to the Fe OEP where the Cl ligand has desorbed, respectively.
\section{Experimental results}\label{sec:exp}
\begin{figure}[tb]
\includegraphics[width=0.38\textwidth]{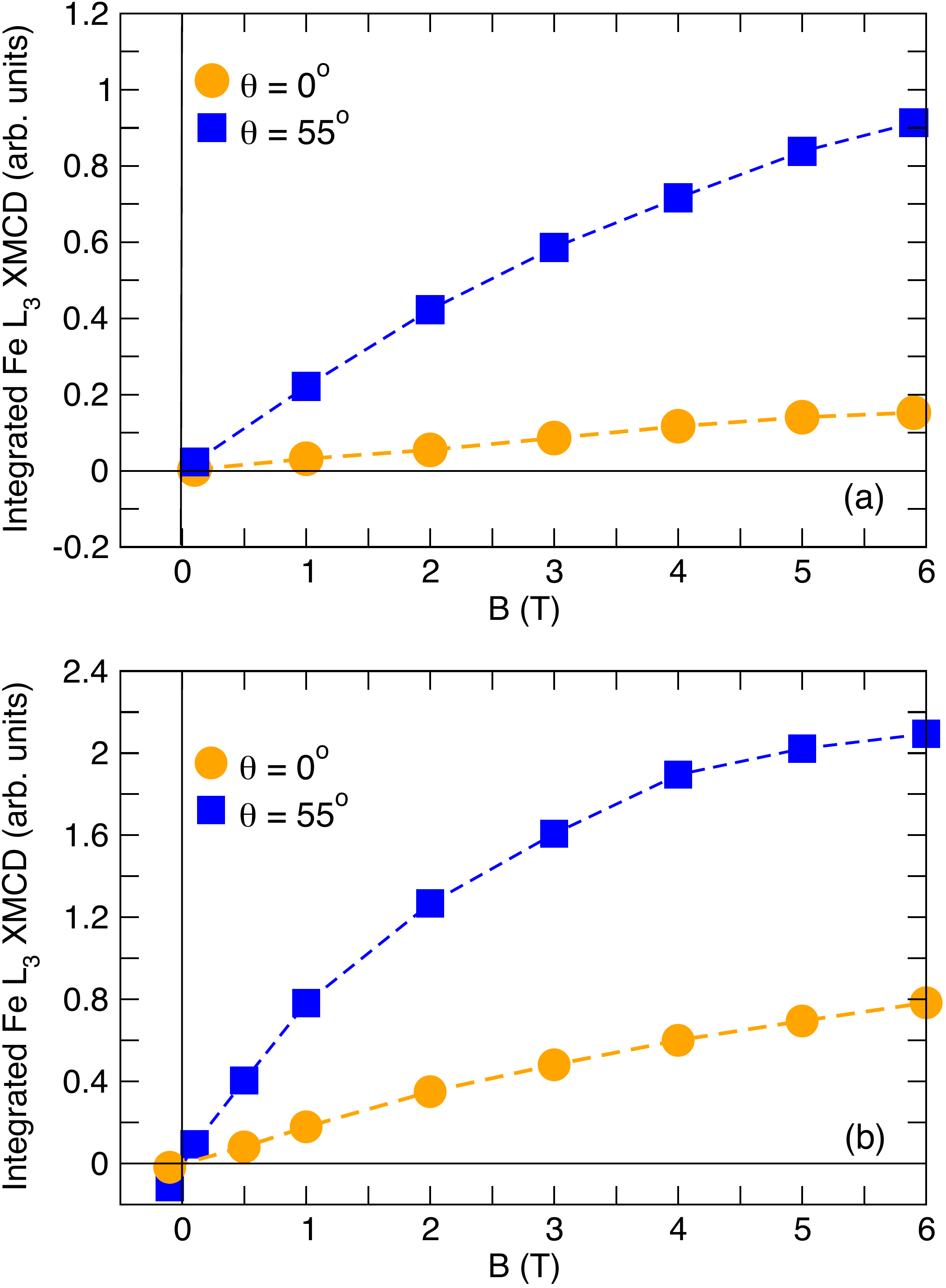}
\caption{(Color online) Integrated Fe $L_3$ XMCD signals of 0.4\,ML of Fe OEP on Cu(001) (a) and on O/Cu(001) (b) at $T =$ 5\,K as a function of applied magnetic field aligned parallel with the $\bf k$ vector of the X rays along $\theta = 0^{\circ}$  and $\theta = 55^{\circ}$ .} 
\label{fig:mag-FeP-Cu}
\end{figure}

\begin{figure*}[bt]
\includegraphics[width=0.85\textwidth]{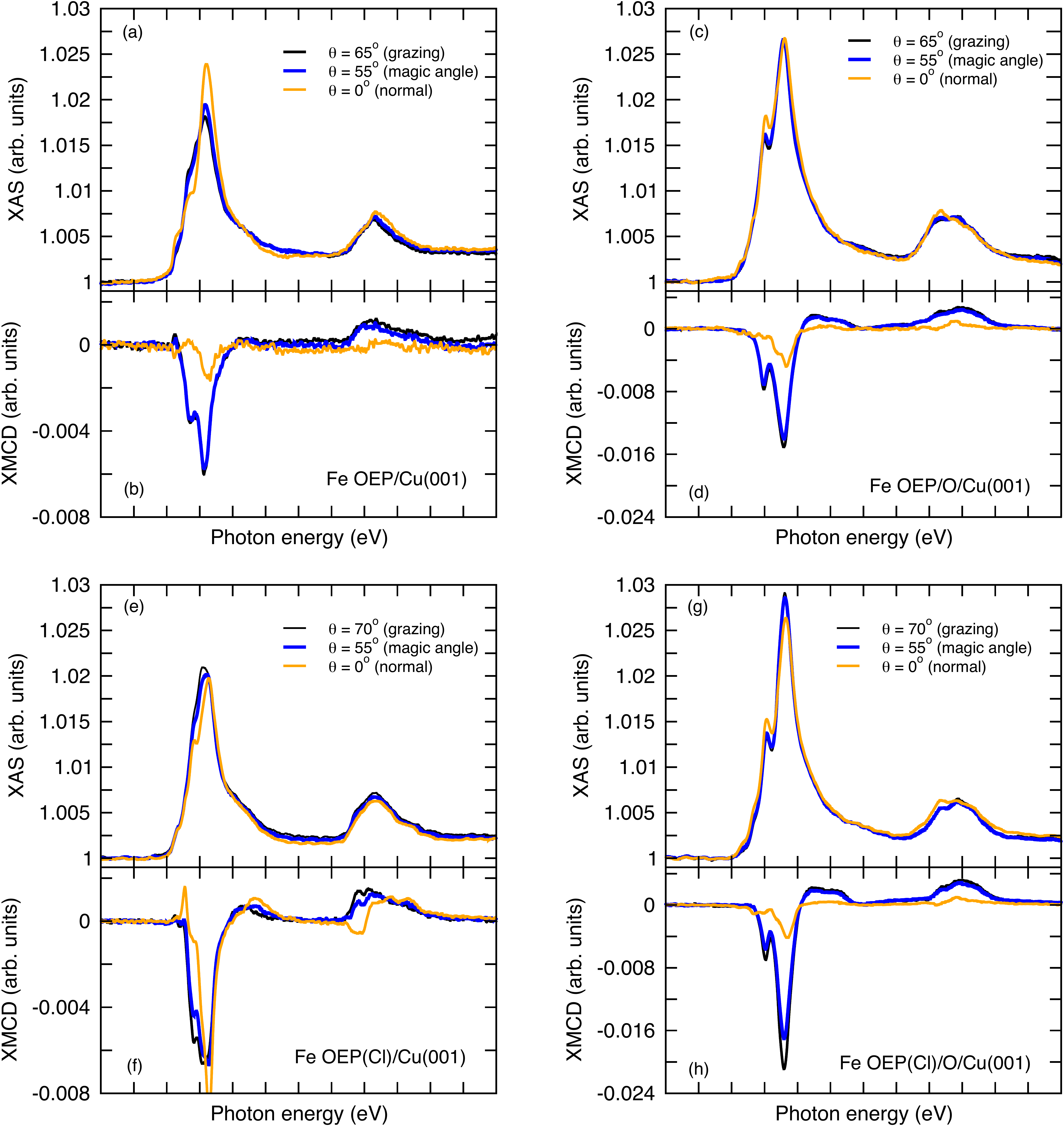}
\caption{(Color online) Fe $L_{2,3}$ XAS ((a), (c), (e), (g)) and XMCD ((b), (d), (f), (h)) signals of 0.4\,ML Fe OEP on Cu(001) and (O)/Cu(001) substrates  (panel (a) to (d)) and Fe OEP Cl (panel (e) to (h)) on substrates of Cu(001) and (O)/Cu(001)  for different photon incidence angle $\theta $ at $T =$ 5\,K ((a)-(d)) and $T =$8\,K ((e)-(h)), respectively. Note the different scales for the XMCD signals.}
\label{fig:XAS}
\end{figure*}

In Fig.\,\ref{fig:mag-FeP-Cu} the integrated Fe $L_3$ XMCD signal of 0.4\,ML Fe OEP on Cu(001) and oxygen-covered Cu(001) is plotted as a function of applied magnetic field along 0$^{\circ}$ and 55$^{\circ}$ at $T = 5$\,K. In both cases the XMCD at 55$^{\circ}$ is significantly higher than the one at 0$^{\circ}$. This can be a result of either an in-plane magnetic anisotropy, a large magnetic dipolar term $\langle T_z\rangle$, or a mixture of both effects. A detailed discussion is found in Sec.\,\ref{sec:aniso}. For Fe OEP on Cu(001) (Fig.\ref{fig:mag-FeP-Cu}(a)), the magnetization curves are almost straight indicating that the Fe ions are magnetically not saturated at $B=6$\,T. The bended curves for Fe OEP on O/Cu(001) (Fig. \ref{fig:mag-FeP-Cu}(b)) on the other hand suggest that magnetic saturation can be reached at 55$^{\circ}$ and $B = 6$\,T for this system.
The Fe $L_{2,3}$ XAS and XMCD signals of 0.4\,ML Fe OEP (Fe OEP (Cl)) on Cu(001) and oxygen-covered Cu(001) are shown in Figs.\,\ref{fig:XAS}(a) to \ref{fig:XAS}(d) (\ref{fig:XAS}(e) to \ref{fig:XAS}(f)) for three different incidence angles at $B = 5.9$\,T ($B = 5.0$\,T) and a temperature of $T = 5$\,K ($T = 8$\,K). The Fe $L_3$ XAS signal of Fe OEP on Cu(001) comprises of a main resonance at 708\,eV mostly visible for normal incidence and three shoulder-like features at 705.7\,eV, 706.6\,eV, and 707.2\,eV. The integrated Fe $L_{2,3}$ XAS signal displays a strong angle-dependent variation even for circularly polarized light. For linearly polarized X rays, a ratio of 1.5 between the integrated Fe $L_{2,3}$ XAS signal (not shown) for 0$^{\circ}$ and 65$^{\circ}$ incidence angle is found indicative for an Fe 3$d$ hole density distribution that is predominantly oriented in the surface plane. The Fe $L_3$ XAS signal of Fe OEP on O/Cu(001) displays two main resonances at about 707.5\,eV and 709\,eV. The overall Fe $L_3$ XAS intensity is shifted by approximately 1.0\,eV to higher photon energy compared to Fe OEP on Cu(001). This shift can be attributed to a charge transfer from the Fe atom in the molecule to the oxygen atom in the substrate. This is an indication that the valence of the Fe atom is trivalent on the oxygen covered surface. Similar trends have been detected for oxygen-covered magnetic surfaces.\cite{Bernien:09} The integrated Fe $L_{2,3}$ XAS signal displays little variation with the incidence angle representing a nearly isotropic 3$d$ hole density distribution. The Fe $L_{2,3}$ XMCD signal of Fe OEP on Cu(001) is smaller than the one on O/Cu(001). The Fe $L_3$ XMCD signal of Fe OEP on Cu(001) (O/Cu(001)) displays a double peak at 706.8\,eV and 707.8\,eV (707.4\,eV and 709\,eV) for 55$^{\circ}$ and 65$^{\circ}$ incidence angle and a resonance at 708.1\,eV (709\,eV) for $\theta = 0^{\circ}$ that is smaller by about a factor 4 for both systems. Since except for the measurements of Fe OEP on O/Cu(001) at $\theta =55^{\circ}$ and $65^{\circ}$ the Fe magnetic moments were not fully aligned by the magnetic field (\emph{cf.} Fig.\,\ref{fig:mag-FeP-Cu}) the strong angle-dependent variation of the XMCD signal can be a result of either an in-plane magnetic anisotropy or a large magnetic dipolar term $\langle T_z\rangle$, i.e., an anisotropy of the spin-density distribution (\emph{cf.} Sec.\,\ref{sec:aniso}). As it will be discussed in section \ref{sec:aniso} the prominent angular dependence is due to a strong angular-dependent $T_z$ contribution and the magnetic anisotropy of the molecules is very small.

The Fe $L_{2,3}$ XAS and XMCD signals of 0.4\,ML Fe OEP(Cl) on Cu(001) (Fig.\,\ref{fig:XAS}(e) and (f)) display an overall similarity to the ones of 0.4\,ML Fe OEP on Cu(001) (Fig.\,\ref{fig:XAS}(a) and (b)) for the spectra measured at grazing incidence and the magic angle with a less pronounced splitting of the double peak feature of the Fe $L_3$ XMCD. On the other hand the XAS and XMCD signal for normal incidence are markedly different. Most prominently the Fe $L_{2,3}$ XMCD of 0.4\,ML Fe OEP(Cl) on Cu(001) is bigger by a factor of about 4 than the one of 0.4\,ML Fe OEP on Cu(001) at normal incidence resulting in an integrated XMCD signal that is basically the same for the different measurement angles. This shows that the electronic and magnetic properties of Fe OEP(Cl) on Cu(001) are different from the ones of 0.4\,ML Fe OEP on Cu(001). Since the STM pictures presented in Fig.\,\ref{fig:STM} show bright and dark molecular centers with a ratio of about 1:3 it can be assumed that part of the molecules have a Cl ion attached to their iron center and others not. In this case the spectra presented in Fig.\,\ref{fig:XAS}(e) and (f) are superpositions of two contributions that cannot be disentangled easily. The Fe $L_{2,3}$ XAS and XMCD signals of 0.4\,ML Fe OEP(Cl) on O/Cu(001) (Fig.\,\ref{fig:XAS}(g) and (h)) present basically the same spectral features and angle dependence than the ones of 0.4\,ML Fe OEP on O/Cu(001) (Fig.\,\ref{fig:XAS}(c) and (d)), showing that the electronic and magnetic properties are comparable even though Cl ions may be present on the surface for the former sample. The effect of Cl attached to the metal center of the molecules will be discussed in Section\,\ref{sec:fep-lig}. 
The results of the sum rule analysis of the spectra shown in Fig.\,\ref{fig:XAS} is presented later on in this manuscript together with the theoretical results. The number of $d$ holes has been assumed to be $n_h =4$ for the molecules on the plain Cu(001) surface, which is in good agreement with the theoretical values obtained from the present DFT calculations ($n_h$ = 3.95). In case of the oxidized surface $n_h =5$ has been used, because the $L_{2,3}$ edge spectra of Fe are  shifted by about 1\,eV to higher energies, which can be assessed as an indication for Fe$^{3+}$.\cite{Bernien:09}
\section{F\lowercase{e}P on O/C\lowercase{u}(001) and C\lowercase{u}(001)} \label{sec:fep}
\subsection{Calculated structural properties}\label{sec:fep-struc}
Supercell calculations have been performed to determine the ground state geometry of the Fe porphyrin molecule on the plain metal and the oxidized surface, where different adsorption sites as well as different orientation relative to the substrate have been taken into account. The results are summarised in Tab.\,\ref{tab:struc}.
\subsubsection{FeP on Cu(001)}\label{sec:fep-struc-o}
\begin{table*}[bht]
\caption{Calculated energy differences $\Delta E$ and Fe-surface distances $d_{\rm Fe-S}$ for FeP  on different adsorption positions of bare Cu(001) and  $\sqrt{2}\times2\sqrt{2}{\rm R}45^{\circ}$ O/Cu(001), with respect to the total energy of the ground state configuration. The values obtained without van der Waals correction are given in brackets. }
\begin{tabular}{c|cccc|cccc}\hline
&\multicolumn{4}{c|}{without O}& \multicolumn{4}{c}{with O}\\\hline
Position\hspace{.4cm}\mbox{} &hollow\hspace{.2cm}\mbox{}&hollow\hspace{.2cm}\mbox{}&bridge\hspace{.2cm}\mbox{}&top\hspace{.2cm}\mbox{}&hollow\hspace{.2cm}\mbox{}& top\hspace{.2cm}\mbox{}&missing\hspace{.2cm}\mbox{}& bridge\hspace{.2cm}\mbox{} \\
 of Fe\hspace{.4cm}\mbox{}&site\hspace{.2cm}\mbox{}&site 45$^{\circ}$\hspace{.2cm}\mbox{}&site\hspace{.2cm}\mbox{}&of Cu\hspace{.2cm}\mbox{}&site\hspace{.2cm}\mbox{}& of O\hspace{.2cm}\mbox{}&row\hspace{.2cm}\mbox{}&site\hspace{.2cm}\mbox{}\\\hline
$\Delta E$ (eV)\hspace{.4cm}\mbox{}&0.0 (0.0) \hspace{.2cm}\mbox{}&0.311 (0.068)\hspace{.2cm}\mbox{}&(5.286)\hspace{.2cm}\mbox{}&(0.485)\hspace{.2cm}\mbox{}&0.566 (0.015)\hspace{.2cm}\mbox{}&    0.580  (0.040)\hspace{.2cm}\mbox{}      &0.0 (0.0)\hspace{.2cm}\mbox{}& 1.840 (0.041)\hspace{.2cm}\mbox{}\\
d$_{\rm Fe-S}$ (\AA)\hspace{.6cm}\mbox{}&2.66 (3.10)\hspace{.2cm}\mbox{} &  2.74 (3.37)\hspace{.2cm}\mbox{}&(3.43)\hspace{.2cm}\mbox{}&(3.11)\hspace{.2cm}\mbox{}&  2.72 (3.48)\hspace{.2cm}\mbox{}& 2.86 (3.50)\hspace{.2cm}\mbox{} &2.69 (3.26)\hspace{.2cm}\mbox{}& 2.87 (3.51)\hspace{.2cm}\mbox{}\\\hline
\end{tabular}
\label{tab:struc}
\end{table*}
\begin{figure}
\includegraphics[width=0.97\columnwidth]{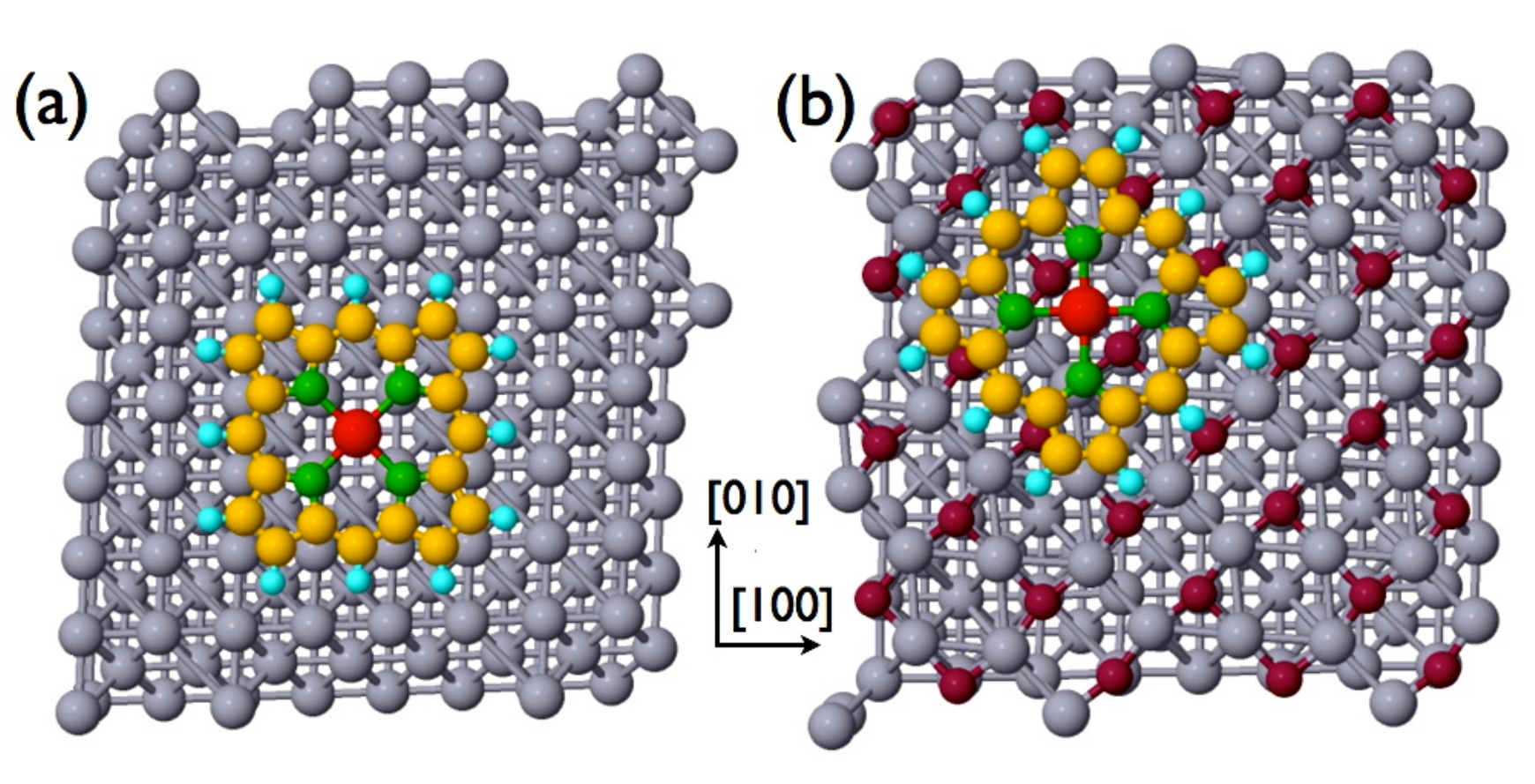}
\caption{(Color online) Schematic view of the ground-state configurations of FeP on Cu(001) (a) and in presence of an intermediate oxygen layer and a  $\sqrt{2}\times2\sqrt{2}\,{\rm R}45^{\circ}$ reconstructed surface (b). Gray spheres mark the Cu atoms of the substrate, purple corresponds to O atoms. The FeP consists of an Fe central atom (red) surrounded by 4 N-atoms (green) and 20 C atoms (yellow). The ethyl groups have been replaced by 12 H atoms (cyan).} 
\label{fig:pos-Fe-OCu}
\end{figure}
In case of a clean Cu(001) substrate the FeP molecule prefers the hollow site position, whereby the four N atoms are located on top of the neighboring Cu atoms of the surface layer, see Fig.\,\ref{fig:pos-Fe-OCu} (a). The distance between the 
central Fe atom and the surface amounts to 3.10\,\AA, if no vdW interaction is considered. Taking into account van der Waals forces in the approximation of Grimme\,\cite{Grimme:06} the distance between molecule and surface is significantly reduced by $\approx 0.4$\,\AA~ to 2.66\,\AA. This bonding distance on the nonmagnetic substrate is even with van der Waals interaction much larger compared to bonding distances obtained for such molecules on FM surfaces\,\cite{Bhandary:12,Lach:12} i.e., FeP or Fe phthalocyanine molecules are chemisorbed on magnetic surfaces.
 This means  that in contrast to a magnetic metal surface the hybridisation between molecule and the bare Cu(001) surface is much weaker. A rotation of the  porphyrin molecule  by 45$^{\circ}$ such that the N atoms are now located between the Cu atoms of the surface layer is connected with an increase of the energy by  0.311\,eV (without van der Waals 0.068\,eV). The  predilection of the N atoms for the Cu atoms of the surface is related to the fact that the hybridization between molecule and surface is mediated by the N-atoms, {\emph cf.} Sec.\,\ref{sec:hybri}. This effect has been previously observed for porphyrin and phthalocyanine molecules on ferromagnetic transition metal  surfaces. \cite{Bhandary:12, Lach:12}
All other positions with Fe on top of a Cu atom or on a bridge position turned out to have significantly higher energies, see Tab.\,\ref{tab:struc}. In case of the bridge position the energy difference to the  hollow site position amounts up to 5\,eV even without van der Waals interaction. For Fe on top of Cu the energy is approximately 0.5\,eV higher than for the ground state, independent of whether the N-atoms are on top of Cu or not. These configurations  will not be considered further. 
\subsubsection{FeP on O/Cu(001)}
The deposition of oxygen on Cu(001) is accompanied by a $\sqrt{2}\times2\sqrt{2}\,{\rm R}45^{\circ}$ reconstruction of the surface and therefore, the surface offers  adsorption sites being different from the ones on Cu(001), see Fig.\,\ref{fig:OCu}(a). However, due to the missing-rows only few positions are available which fit to the 4-fold symmetry of the molecule. Since, O is attractive for Fe in addition to the previously investigated adsorption sites, configurations with Fe on top of O and on the missing-row have been taken into account.  From the calculations it turned out that  the most stable structure is indeed the missing-row position, i.e., the Fe atom sits directly over the missing-row with the 4 N atoms on top of Cu atoms, see Fig.\,\ref{fig:pos-Fe-OCu} (b). Here, the Fe atom has two O neighbors in (-110) direction. This leads to a tiny deformation of the N cage, i.e. the angle between Fe and the two N atoms with O amounts to 89.4$^{\circ}$. 
The hollow site position, which was the favorable position in case of the plain Cu(001) surface, is without van der Waals interaction only  15\,meV  higher in energy. Configurations with Fe on top of an O atom or on a bridge-like position (Fe in the middle between Cu and O) can be found about 40\,meV (without van der Waals forces) above the ground state configuration, whereby for Fe on top of O only three of the four N atoms have a Cu counterpart to hybridize with. In case of the less symmetric bridge-like position none of the N-atoms has the chance to interact directly with a Cu atom, instead two of the C atoms of  each pyrrole unit sit on top of the Cu atoms. However, the relatively small energy differences between the different positions suggest that all four  positions may be partially  occupied at finite temperatures. 
The picture drastically changes if van der Waals forces are included. Although the ground state configuration remains the same and  also the energetic sequence  of the adsorption positions is the same,  the missing-row position becomes more  stable compared to the hollow site, top of oxygen, and bridge-like positions, see Tab.\,\ref{tab:struc}. The missing-row position is now 0.566\,eV lower in energy compared to the hollow site position. Obviously, the long-range van der Waals forces drift apart the energies of the different adsorption sites, which is related to the reduction of the molecule-surface distance and the related stronger surface-molecule interaction. 
The results for the 45$^{\circ}$-rotated positions are not given in Tab.\,\ref{tab:struc}, because they are the  less favorable configuration on each adsorption site. For example rotating the molecule on the missing-row position by 45$^{\circ}$, such that two of the N-atoms end up sitting on top of O, costs approximately 0.7\,eV without van der Waals correction, and the magnetic moment vanishes.
Without van der Waals forces the distance between molecule and surface is so large (> 3.2\,\AA) that the interaction with  the surface is rather weak, whereas with van der Waals forces the hybridization between molecule and surface increases. The distance between Fe atom and oxygen is now reduced to 2.65\,\AA~for the ground state configuration. As a consequence, asymmetric configurations will be rather unlikely, e.g., the configuration with Fe  on top of O is now 0.580\,eV higher in energy because only three N atoms have the possibility to hybridize with Cu atoms of the underlying surface,  whereas the fourth N-atom is located on the missing-row position, see Fig.\,\ref{fig:pos-Fe-OCu} (b). This clearly demonstrates the importance of van der Waals corrections in this type of systems. 
However, in experiment other than missing-row position may be partially occupied for larger molecule concentrations and in case of low growth temperatures, where the mobility of the molecules is limited and they may stick to unfavorable positions. Therefore, in the following we discuss the ground state configuration and the configuration being closest in energy to the ground state.
\begin{figure*}
\begin{center}
\includegraphics[width=0.9\textwidth]{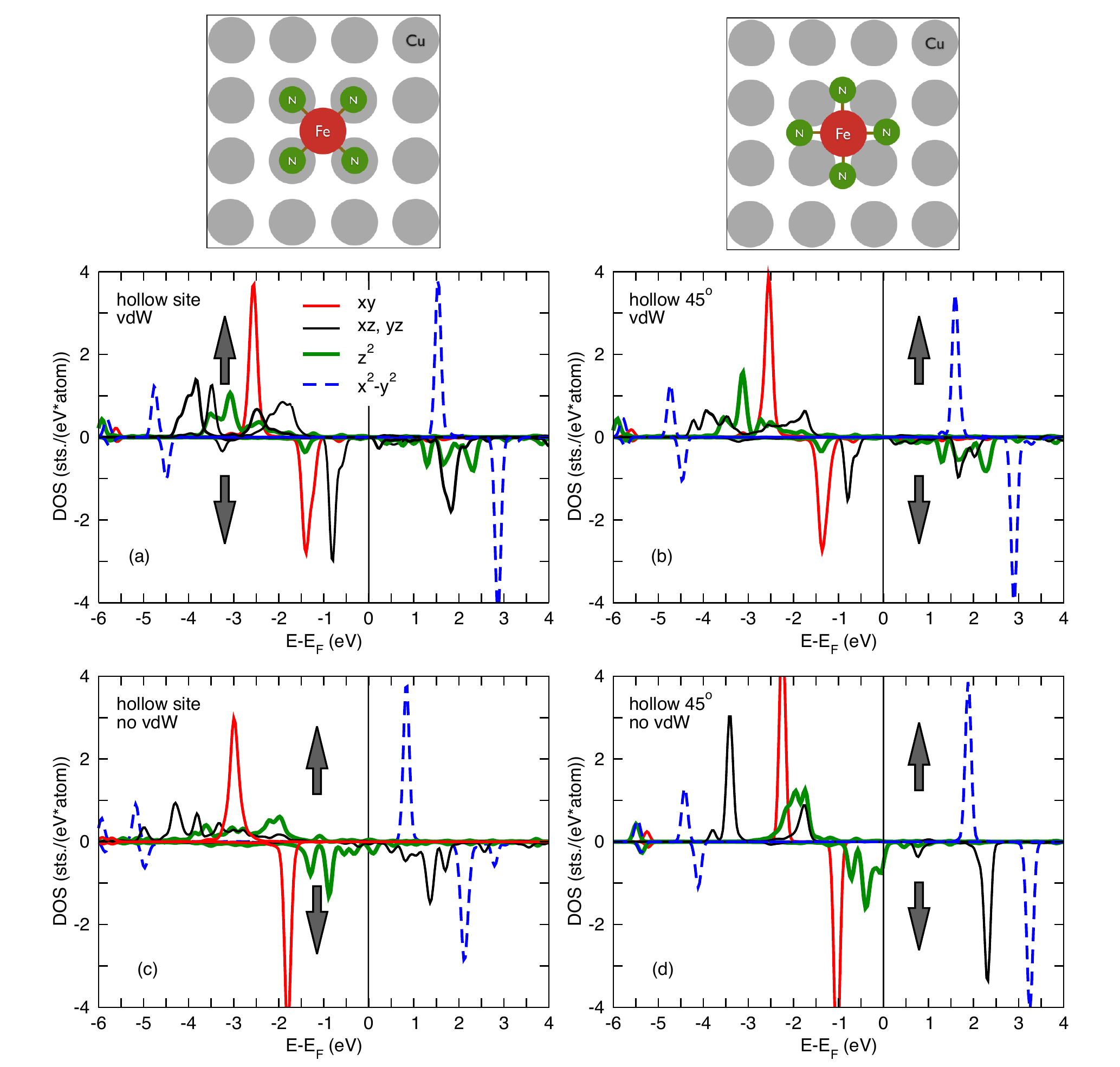}
\caption{(Color online) Calculated density of states of Fe $d$ orbitals in FeP/Cu(001). Arrows denote the spin-up and spin-down configuration, respectively. The Fe atom is adsorbed on the hollow site position with N on top of Cu (a) and (c). In case of (b) and (d) the molecule is rotated by 45$^{\circ}$. The results obtained from calculations with (without) van der Waals forces in the approximation of Grimme\,\cite{Grimme:06} are given in the top row (bottom row). Differently colored lines denote different $d$ orbitals as indicated in the legend in (a).}  
\label{fig:dos-Fe-Cu}
\end{center}
\end{figure*}
 It should be noted that the inclusion of van der Waals forces is responsible for a roughening of the surface layer. The surface, which had previously been  mainly  flat, shows now a distinct buckling, i.e., the height of the O atoms varies by 0.78\,\AA. O atoms under the molecule move inwards whereas the other stick out of the surface. The same effect is observed on the plain Cu(001) surface, but it is much less expressed.
\begin{figure*}
\includegraphics[width=0.9\textwidth]{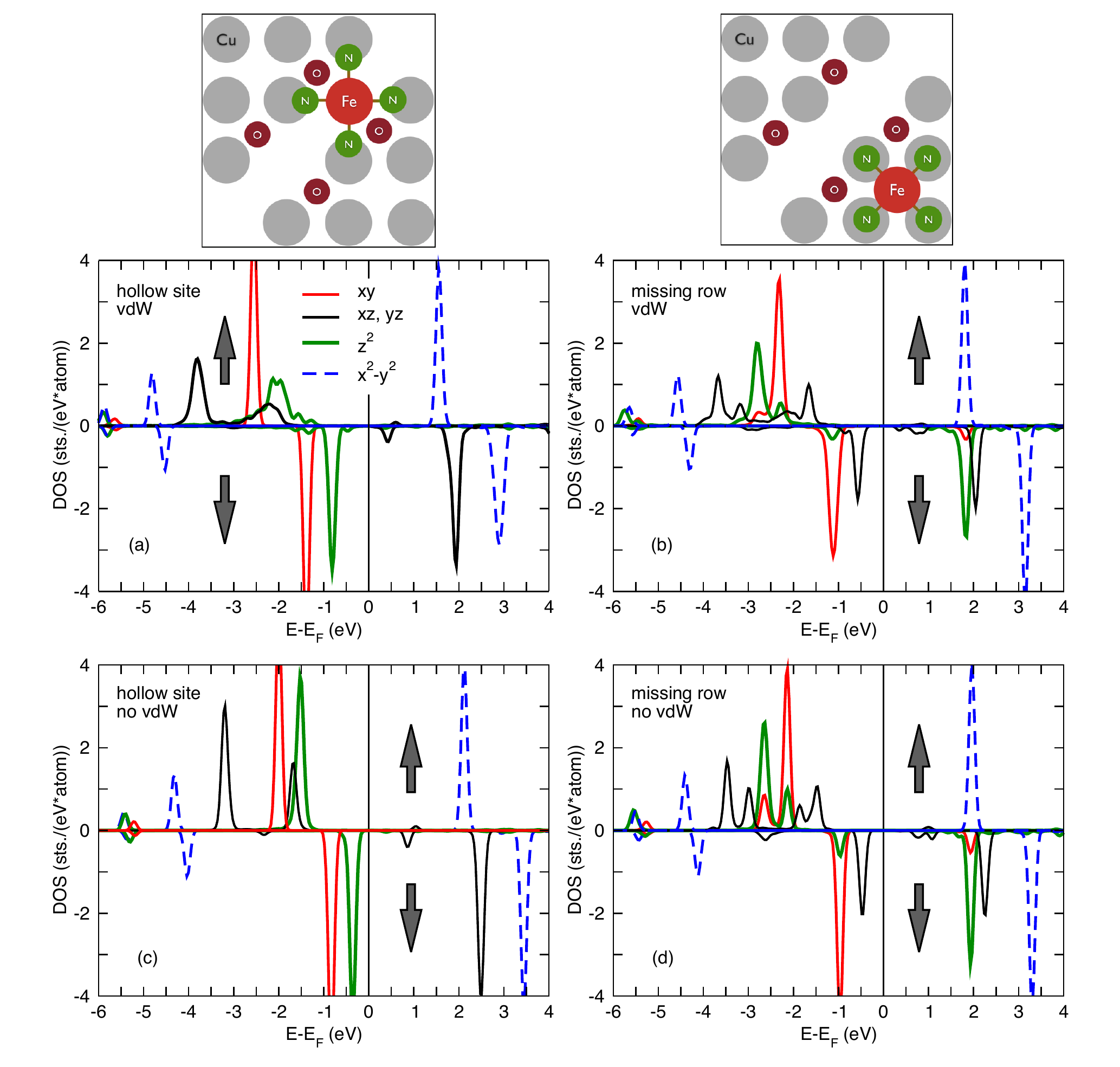}
\caption{(Color online) Calculated \dos{} of Fe in FeP/ $\sqrt{2}\times2\sqrt{2}\,{\rm R}45^{\circ}$O/Cu(001). Arrows denote the spin-up and spin-down configuration, respectively. The Fe atom is placed on the hollow site ((a), (c)) and the missing-row position ((b) and (d)) with the N-atoms on top of Cu. Results including van der Waals (vdW) forces are shown in the top row, results without van der Waals interaction are given in the bottom row. Differently colored lines represent different $d$ orbitals as indicated in the legend in (a).} 
\label{fig:dos-Fe-OCu}
\end{figure*}
\subsection{Calculated magnetic properties}\label{sec:mag}
From here on we focus only on the two most stable configurations, i.e., the ground state and state which is closest in energy to the ground state.
\subsubsection{FeP on Cu(001)}
The FeP molecule on Cu(001) carries a spin moment of 2$\,\mu_{\rm B}$ and is in an intermediate spin state $S=1$ for all possible adsorption sites, which agrees with our  results from the sum-rule analysis.  For a  more direct comparison of the magnetic moments with  the effective magnetic moments from the sum-rule analysis the dipolar term has to be included, see discussion in Sec.\,\ref{sec:aniso}. The $S=1$ spin-state turns out to be very stable and is not changed when van der Waals forces are included. If in addition a high-spin (4\,$\mu_{\rm B}$) state can be stabilized it has always much higher energy, e.g. for the ground state configuration the high-spin state lies without van der Waals interaction $\approx$0.5\,eV above the intermediate spin-state.
The \dos{} of the Fe $d$ orbitals of FeP on the hollow site position can be found in Fig.\,\ref{fig:dos-Fe-Cu}. For the ground state hollow site position (N on top of Cu)  and the $45^{\circ}$ rotated molecule (N between Cu) a $(d_{xy})^2$, $(d_{\pi})^3$, $(d_{z^2})^1$ configuration appears if van der Waals forces are included. This corresponds to recent findings for the gas phase FeP molecule\,\cite{Panchmatia:13}. In contrast to that, without van der Waals interaction the  $d_{z^2}$ orbital is mostly filled, such that the occupancy corresponds to  $(d_{xy})^2$, $(d_{\pi})^2$, $(d_{z^2})^2$, see Fig.\,\ref{fig:dos-Fe-Cu}. Besides the change in occupation of the $z$-oriented orbitals the density of states with and without van der Waals interaction are very similar.  In contrast to the gas phase FeP molecule the \dos{} is broadened and no clear HOMO-LUMO gap can be observed. The ligand field splitting which can be obtained from the energy difference between the $d_{xy}$ and $d_{x^2-y^2}$ orbital amounts to $\approx$ 1\,eV in all cases, when the artificial splitting due to the Hubbard-like $U-J$ term has been subtracted. 
\subsubsection{FeP on O/Cu(001)} 
As discussed in Sec.\,\ref{sec:fep-struc-o} the adsorption site of the molecule is different on $\sqrt{2}\times2\sqrt{2}\,{\rm R}45^{\circ}$O/Cu(001) compared to Cu(001), however this has just as little influence on the magnetic structure resulting in a slightly deformed Fe-N cage. The magnetic configuration remains unchanged, i.e., the Fe atom is in an intermediate spin state and carries a spin moment of 2$\,\mu_{\rm B}$. For the Fe atom on the missing-row position, the occupation of the orbitals is the same as for the previous case ($(d_{xy})^2$, $(d_{\pi})^3$, $(d_{z^2})^1$), independent from whether van der Waals interaction is included or not, see Fig.\,\ref{fig:dos-Fe-OCu}(b), (d). The HOMO-LUMO gap amounts to 0.9\,eV, which is only slightly smaller than the value obtained for the gas phase molecule\,\cite{Panchmatia:13}. In contrast to the observations on the missing-row position, for molecules on the hollow site position  the $d_{z^2}$ orbital is completely filled, see Fig.\,\ref{fig:dos-Fe-OCu}(a) and (c). Interestingly, the system has -- at least in the unoccupied part --  narrow energy bands, i.e., sharp peaks in the \dos, which is different from the observations on the plain Cu(001) surface. Here, the shape of the \dos{} is more alike to the gas phase molecule\,\cite{Panchmatia:13}. Only the occupied $d_{\pi}$ and $d_{z^2}$ orbitals are  somewhat broadened compared to the free Fe porphyrin, which stems from  interaction with the surface, see discussion in the next section. This broadening is not observed for the hollow site position without van der Waals interaction (Fig.\,\ref{fig:dos-Fe-OCu}(c)), which is explained by the huge molecule-surface distance of 3.48\,\AA. 
\subsubsection{Hybridization and coupling to the surface layer}\label{sec:hybri}
\begin{figure}[bt]
 \includegraphics[width=0.4\textwidth]{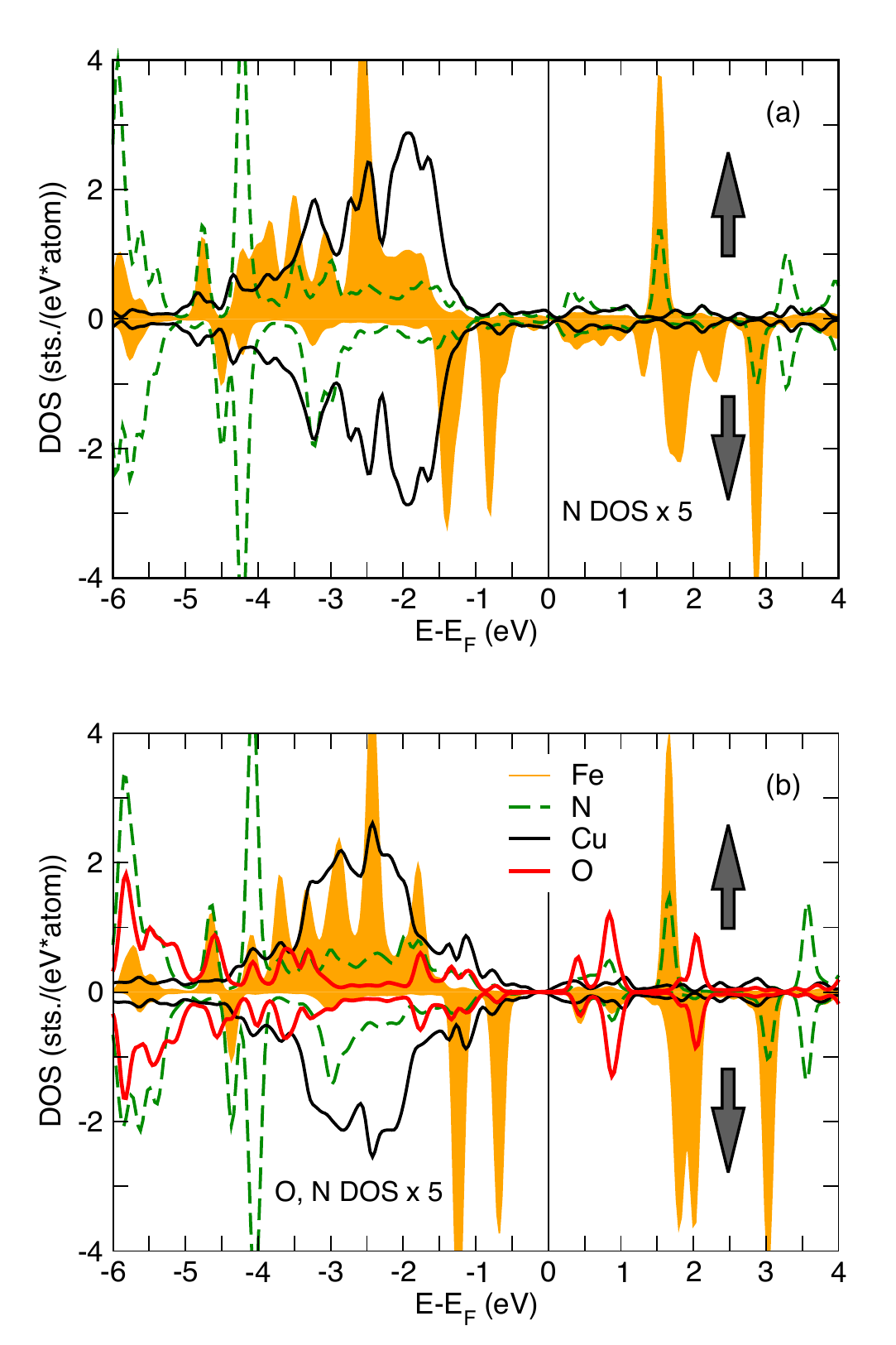}
\caption{(Color online) Calculated density of states of Cu and O from the surface layer and \dos{} of N atoms and  Fe $d$ orbitals from the FeP molecule. (a) FeP on the hollow site position on the  plain Cu(001) surface and (b) on missing-row position of  the oxidized surface, respectively. Calculations are performed with van der Waals interaction.\cite{Grimme:06} Arrows denote the spin-up and spin-down configuration, respectively.}  
\label{fig:dos-surf-mol}
\end{figure}

So far the electronic, magnetic and structural properties of the molecules themselves have been studied, here we focus on  the interaction between molecule and surface. It has been shown in Sec.\,\ref{sec:fep-struc} that the distance between molecule and surface even with van der Waals correction is about 2.7\,\AA, thus the hybridization effects with the surface are weaker compared to adsorption on magnetic substrates.\cite{Bhandary:12}
Without oxygen the interaction between Fe and Cu atoms is small. On one hand small hybridization can be assumed from peaks around 2-3\,eV below the Fermi level in the majority-spin channel, which appear for both systems, but no significant hybridization could be observed for the minority-spins, see Fig.\,\ref{fig:dos-surf-mol}(a). This suggests that the direct interaction between Cu atoms and Fe atom plays only a minor role. On the other hand as can be seen from Fig.\,\ref{fig:dos-surf-mol}(a), without oxygen interlayer the positions of the  N orbitals overlap with Cu and  Fe orbitals, i.e., the interaction between molecule and surface is also mediated by the N atoms.  This matches with the fact that the preferred orientation of the molecule on the surface is with the N-atoms on top of Cu, {\emph cf.} Sec.\,\ref{sec:fep-struc}. This type of indirect coupling has been observed on magnetic surfaces, e.g. for FeP on Co(001).\cite{Bhandary:12} The charge of the Fe atom and the four N atoms is reduced by 0.01e compared to the gas phase and the total charge of the Cu atoms in the vicinity of the molecule is increased compared to the free Cu(001) surface.

In case of the oxygen-covered surface  the situation becomes more complex, see Fig.\,\ref{fig:dos-surf-mol}(b).  Here, the hybridization between Cu and Fe $d$ electrons seems to be  somewhat larger compared to the plain Cu surface, especially in the majority channel the Cu d state positions coincide with the Fe $d$ orbital.  This means in contrast to the previous case the coupling is not dominated by the indirect coupling of Fe to Cu via N.  Instead the Fe atom hybridizes with the neighboring O atoms, for example in the minority channel at 1\,eV below and 2\,eV above the Fermi level. Similar as before the charge of the Fe atom is slightly reduced compared to the charge of the Fe atom in the gas phase molecule. Here, a small charge transfer from the Fe atom to the O and Cu atoms under and next to the molecule is likely, because the oxygen charge is larger by 0.014e  as without the molecule. The same holds for the Cu charge. The N charge remains unchanged. Although, the changes of the individual charges are small, they are in line with the experimental findings for Fe OEP on the O covered surface, {\emph cf.} Sec.\,\ref{sec:exp}. However, the calculated charge variation is much too small to indicate the existence of  trivalent Fe.
The fact that a charge transfer occurs from Fe  to O indicates that  the Fe atom directly hybridizes with the surface. Similar effects have been observed for FeP on O/Co(001)\,\cite{Bernien:09}. However, there the driving force is the magnetic exchange coupling which does not exist here. However,  we observe a
hybridization between O and N orbitals ($s$ and $p$ type) 4 eV and more below the Fermi energy, see Fig.\,\ref{fig:dos-surf-mol}, which means that  an indirect binding exists at least partially.

Although there occurs a small charge transfer between molecule and surface and hybridization between the orbitals of the molecule and the ones from the surface exist, the  situation is completely different from the scenario on the magnetic substrates, where a magnetic coupling between molecule and surface exists.\cite{Bernien:09}
\section{Influence of O and C\lowercase{l} ligands} \label{sec:fep-lig}
\begin{figure}[tb]
\includegraphics[width=0.35\textwidth]{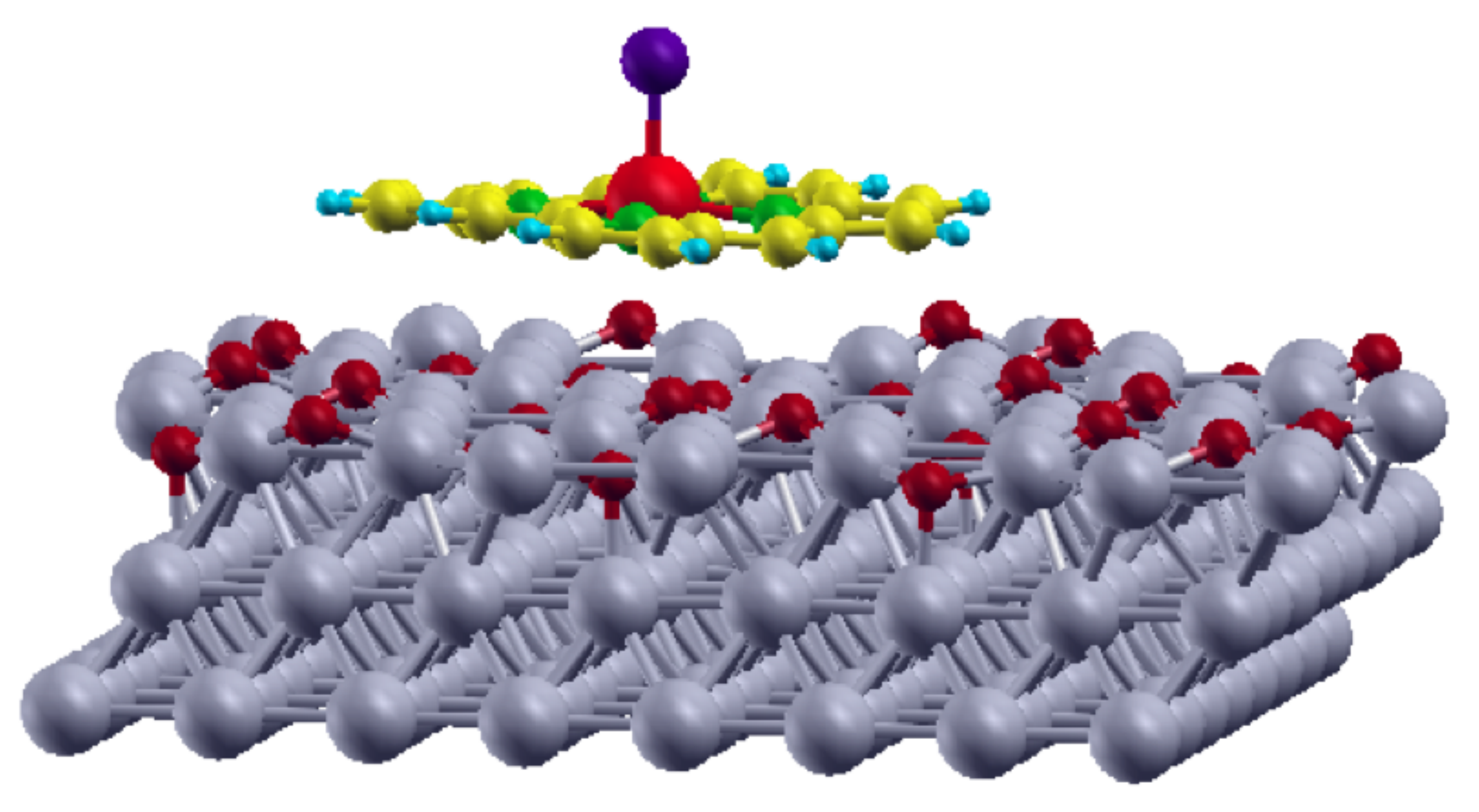}
\caption{(Color online) Sketch of optimized structure of FeP with Cl ligand on  $\sqrt{2}\times2\sqrt{2}\,{\rm R}45^{\circ}$O/Cu(001).} 
\label{fig:FeP-Cl}
\end{figure}
\begin{figure}[b]
\includegraphics[width=0.9\columnwidth]{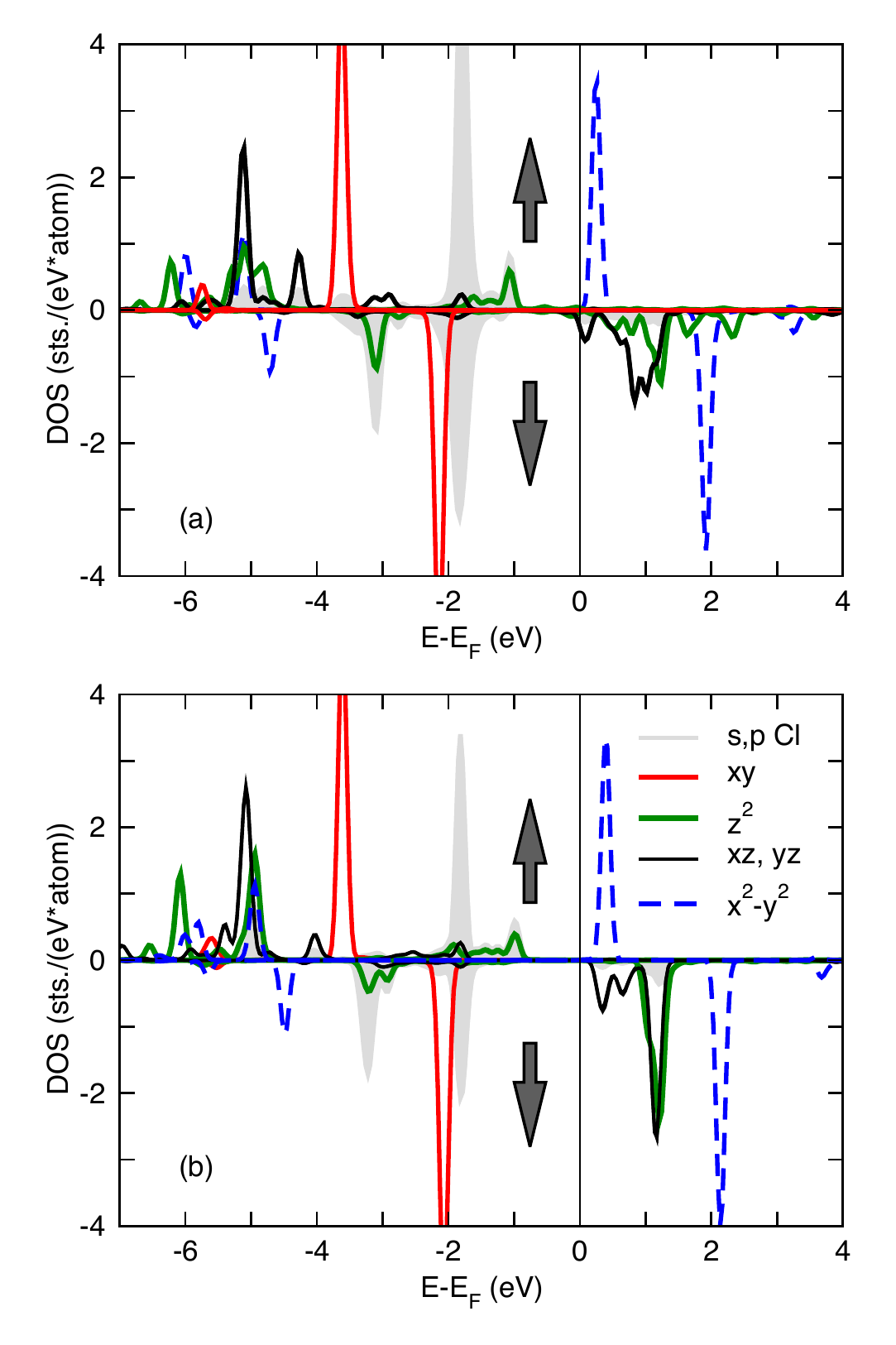}
\caption{(Color online) Density of states of Fe $d$-states and Cl (gray) in FeP(Cl) on Cu(001) (a) and  on O/Cu(001) (b) with van der Waals interaction. The position of the molecule corresponds to the respective ground state configuration, see Sec.\,\ref{sec:fep-struc}. Arrows denote the spin-up and spin-down configuration, respectively.}
\label{fig:dos-Fe-Cu-Cl}
\end{figure}
Fe porphyrin molecules are delivered with stabilizing ligands such as Cl or pyridine (C$_5$H$_5$N). Although the ligands should be removed during the preparation process, especially Cl atoms may remain in the system and stick to the FeP molecules. This is suggested by STM  and XAS measurements of Fe OEP(Cl) on Cu(001) and discussed in Sec.\,\ref{sec:exp}. In case of an oxygen interlayer there may also be some residual oxygen gas remaining in the chamber. Therefore, we have investigated the influence of these two types of ligands attached to the FeP molecule on the oxidized surface, respectively. In case of Cu(001)  only Cl ligands have been considered, because no residual oxygen gas is expected in this case. Single Cl or O atoms have been attached to  the Fe-atom at the remote side from the surface, see Fig.\,\ref{fig:FeP-Cl}. 
Here, the discussion is restricted to calculations including van der Waals forces. 
\begin{table}[hbt]
\caption{Calculated Fe-surface distances $d_{\rm Fe-surface}$  and  bond length of  Fe and ligand atom $d_{\rm Fe-ligand}$  for FeP on  missing-row position  $\sqrt{2}\times2\sqrt{2}{\rm R}45^{\circ}$ O/Cu(001) (named FeP/O/Cu(001) in the table) and on the hollow site position on Cu(001).  In addition the Fe-N distance $d_{\rm Fe-N}$ is given. In all calculations van der Waals forces employing Grimme correction  have been taken into account.}
\begin{tabular}{lccc}\\\hline\hline
System\vspace{.1cm}&FeP/Cu(001) &\multicolumn{2}{c}{FeP/O/Cu(001)}\\\hline
Ligand\vspace{.1cm}& Cl&Cl  &O \\\hline
$d_{\rm Fe-surface}\vspace{.1cm}$ (\AA)&2.64&\hspace{.5cm}3.08\hspace{.5cm} &3.16\\
$d_{\rm Fe-ligand}\vspace{.1cm}$ (\AA)&2.29&2.26&1.68\\
$d_{\rm Fe-N}\vspace{.1cm}$ (\AA)&2.04&2.02&2.09\\\hline\hline
\end{tabular}
\label{tab:struc-lig}
\end{table}%

Both types of ligands lead to a wide difference in electronic and magnetic structure compared to the ideal FeP molecule on Cu(001) and O/Cu(001).
Chlorine ligands provoke a destabilization of the molecule-surface hybrid system and lead to a concomitant increase of the distance between surface and FeP molecule by 0.43\,\AA~ for FeP/O/Cu(001), see Table\,\ref{tab:struc-lig}. No such destabilization could be observed on the plain Cu(001) surface, because it is related to the electrostatic repulsion between Cl and O atoms. On Cu(001) the distance between surface and molecule remains unchanged compared to the ligand-free case. Due to the relatively large distance  to the surface, the hybridization with the surface atoms is comparable to the one without Cl. In contrast to that, the distance between the Cl and Fe atom amounts to 2.26\,\AA~(2.29\,\AA~ on Cu(001)),  which is of the same order as the bond lengths in the molecule itself.  Consequently, the $d$ states of Fe which stick out of the molecular plane ($d_{z^2},\, d_{\pi}$) hybridize with the Cl $p$ states, and the unoccupied $d$ states are shifted  closer to the Fermi level, see Fig.\ref{fig:dos-Fe-Cu-Cl}.  
Due to the weak coupling to the surface, the magnetic and electronic structure of the molecule is strongly influenced by the ligand. If Cl is attached to the Fe atom, the spin moment  increases  from 1.99 to 2.62\,$\mu_{\rm B}$ for the molecule on the oxidized surface and from 1.93  to 2.50\,$\mu_{\rm B}$ for the molecule on the pure metal surface. The electronic configuration is close to $(d_{xy})^2$,$(d_{\pi})^2$,$(d_{z^2})^1$.  The main difference is that the partially occupied $d_{\pi}$ orbital close to $E_{\rm F}$ is now unoccupied. However, the $d_{x^2-y^2}$ orbital is partially occupied such that the ratio of spin-up and spin-down charge density is about 2.5 and a clear decision whether the system is in a Fe$^{2+}$ or Fe$^{3+}$ configuration is not possible. The size of the HOMO-LUMO gap is unaffected by the Cl ligand, but the splitting between the unoccupied $d$ \dos{} becomes larger, being 1.63\,eV  instead of 1.38\,eV without the extra ligand, see Fig.\,\ref{fig:dos-Fe-Cu-Cl}. In the experiment this would express itself in a broadening of the $L_3$ XAS. Furthermore, the pre-edge features are expected to become smaller, because the  majority $d_{x^2-y^2}$ orbital forms now the LUMO and has the same energy as the minority $d_{\pi}$ orbitals, see discussion in Sec.\,\ref{sec:mag}. 

\begin{figure}
\includegraphics[width=0.9\columnwidth]{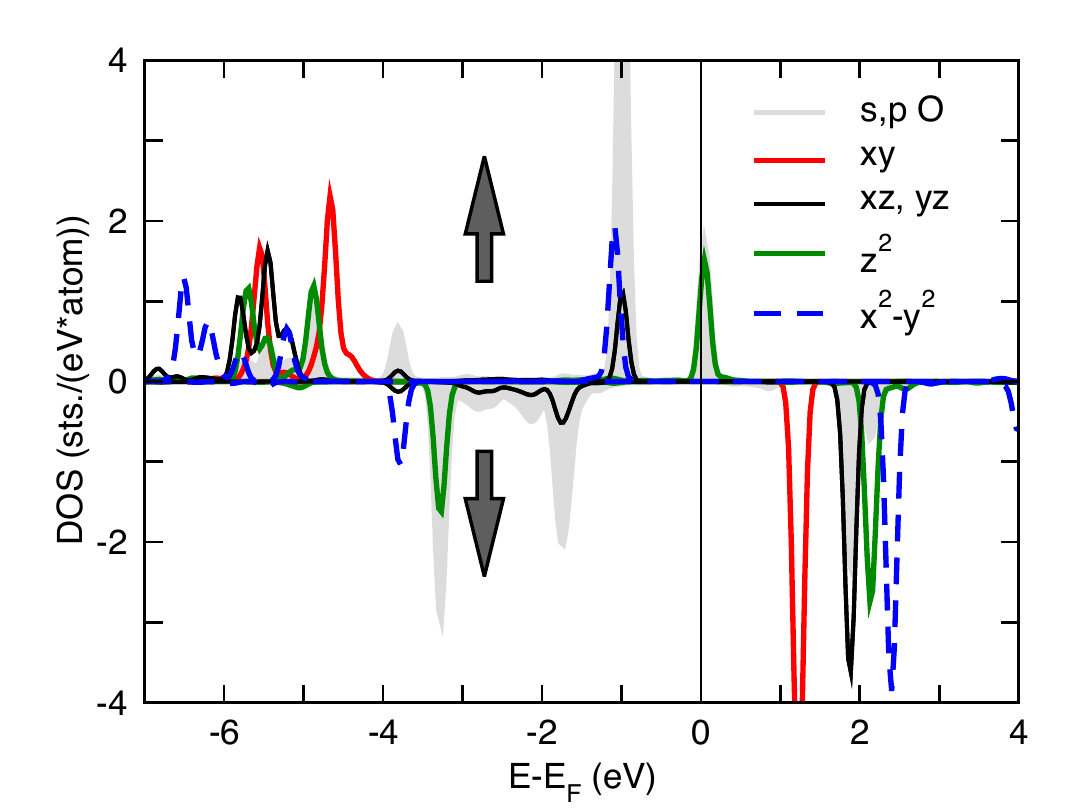}
\caption{(Color online)Calculated density of states of Fe and O (gray) in (FeP(O)) on $\sqrt{2}\times2\sqrt{2}\,{\rm R}45^{\circ}$O/Cu(001) including van der Waals interaction. Arrows denote the spin-up and spin-down configuration, respectively.} 
\label{fig:dos-Fe-Cu-O}
\end{figure}

Assuming that an O atom is attached to the molecule (similar to the case of Cl Fig.\,\ref{fig:FeP-Cl}), or more precisely to the Fe atom, the scenario is very different. On the one hand the structural changes are similar to the previous case, i.e., the bonding strength between surface and molecule decreases due to the ligand. In the presence of an O ligand the  molecule-surface distance amounts to 3.16\,\AA,  whereas the bond length between Fe-atom and ligand is only 1.68\,\AA.  The small bond length is comparable to the findings in bulk FeO, in which the Fe-O distance is about 1.62\,\AA.\cite{Ram:96}   On the other hand the oxygen atom influences the magnetic properties of the molecule much more than the above discussed Cl atom. Here, the spin moment of the Fe atom is 3.50\,$\mu_{\rm B}$, {\em cf.} Tab.\,\ref{tab:magn}. In contrast to the previous case  the Fe atom is now in the $S=2$-like spin-state with $n^{\uparrow}/n^{\downarrow} = 4.66/1.20$, i.e., only one spin-down (mostly $d_{z^2}$ character) orbital remains occupied, see Fig.\,\ref{fig:dos-Fe-Cu-O}.  The small bond length between Fe and O as well as the tendency of Fe to oxidize lead to strong hybridization effects between molecule and ligand, i.e., of  O $p$ orbitals (gray shaded in Fig.\,\ref{fig:dos-Fe-Cu-O}) with  Fe $d_{z^2}$ and $d_{\pi}$ orbitals.
Other than for Cl ligands O strongly influences the HOMO-LUMO gap. A $p$-$d$ hybrid orbital appears directly at $E_{\rm F}$ in the spin-up channel, whereas the large gap in the spin-down channel remains nearly unchanged.
Summarizing, the binding of  Cl or O ligands  to the Fe center of the porphyrin molecule stabilizes the high-spin state. The occurrence of the high-spin state goes along with an increase of the Fe-N distance. In contrast to the previously discussed systems without ligands here the Fe-N bond length lies between 2.02 to 2.09\,\AA, see Tab.\,\ref{tab:struc-lig}.  These observations agree with findings for Fe porphyrin molecules in the gas phase\,\cite{Liao:02} and on graphene\,\cite{Bhandary:11}.

\subsection{Comparison between experiment and theory}
\subsubsection{XAS vs. DOS}
\begin{figure}[tb]
\includegraphics[width=0.485\textwidth]{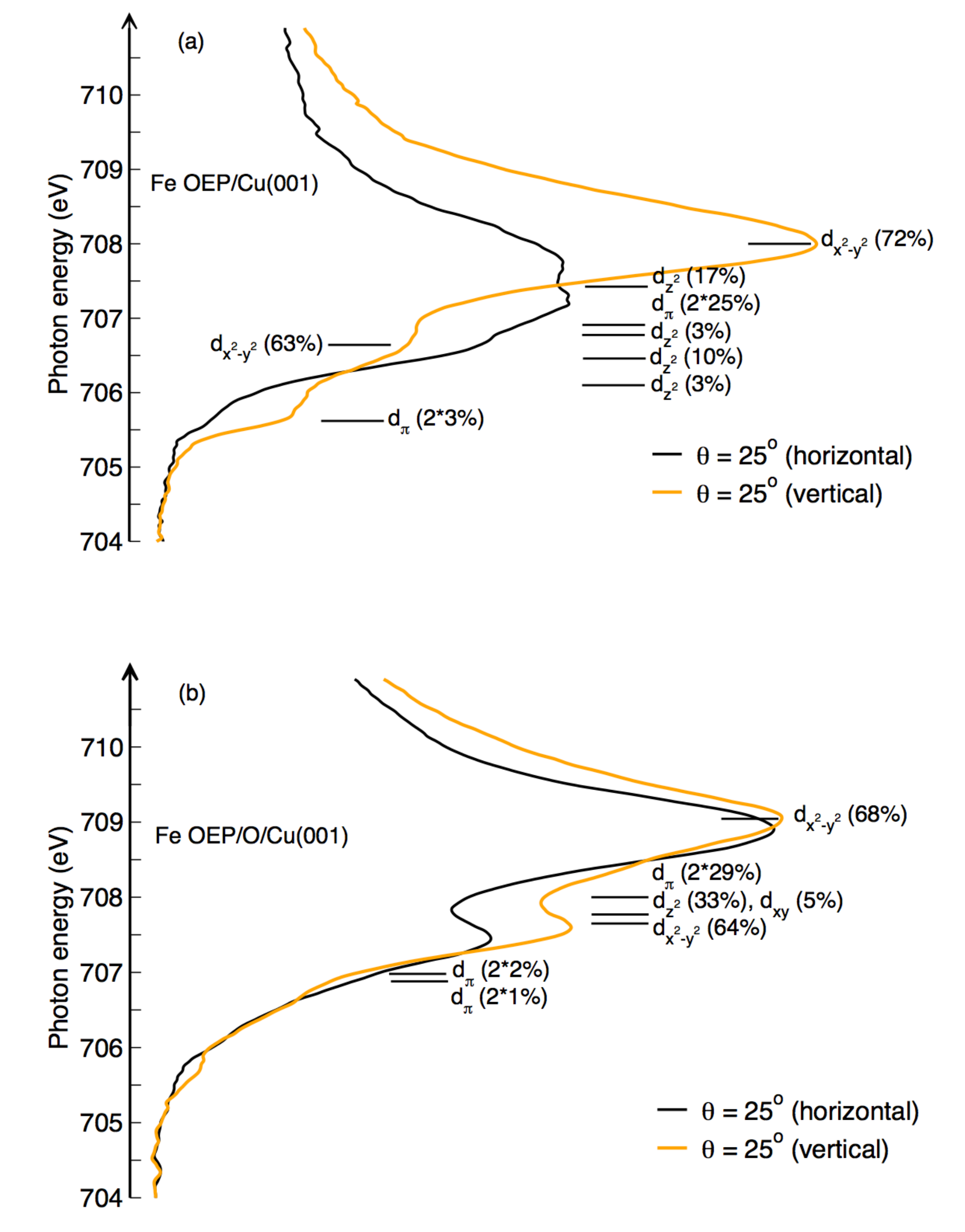}\\
\caption{(Color online) XAS of Fe $L_3$ edge of Fe OEP on Cu(001) (a) and  $\sqrt{2}\times2\sqrt{2}\,{\rm R}45^{\circ}$O/Cu(001) (b) measured with linear polarized light. The calculated energy levels of the unoccupied Fe 3$d$ orbitals are shown as horizontal lines. The percentage of the $d$ character is given in brackets. The percentage of the d character is given in brackets. Black lines mark the XAS with  electric field mostly oriented perpendicular to the surface ($\theta=25^{\circ}$ horizontal) and the bright line denotes the XAS obtained for $\theta = 25^{\circ}$ (vertical), where the electric field vector is parallel to the surface. } \label{fig:xas-dos-pyr}
\end{figure}

In order to compare the theoretical results to X-ray absorption experiments the sequence of the unoccupied $d$ states can be related to the features of the XAS at the Fe $L_3$ edge obtained from linear polarized light. Different parts of the \dos{} will be visible for different incidence angles $\theta$ of the photon beam and polarization orientations. For normal incidence $d$ orbitals in the ($x-y$) plane are scanned and in case of grazing incidence mostly orbitals with $z$ components contribute, see Fig.\,\ref{fig:beam}. In the following we focus on the Fe $L_3$ edge only  and for a  discussion  the energy levels of the unoccupied $d$ states are included in the graph, see Fig.\,\ref{fig:xas-dos-pyr}.

Although we are aware of the fact that a direct comparison of \dos{}~ and XAS is difficult because of missing core-hole effects  in theory, it gives at least information of the character of the XAS peaks. 
The XAS of Fe OEP  on Cu(001) shows a broad Fe $L_3$ edge signal for grazing ($\theta = 25^{\circ}$ (horizontal)) incidence. This  corresponds to the broadened $d_{\pi}$ and $d_{z^2}$ orbitals which we obtained from the present DFT calculations, see Fig.\,\ref{fig:xas-dos-pyr}(a).  In case of normal incidence two peaks occur at  706.8 and  708\,eV. This  is in reasonable agreement with the separation of the two $d_{x^2-y^2}$ orbitals which is 1.33\,eV. The pre-peak feature in Fig.\,\ref{fig:xas-dos-pyr} is more pronounced for normal incidence, i.e., the electrons are excited to $d$ orbitals which are (partially) in-plane oriented. This corresponds to the $d_{\pi}$ states which form the LUMO orbital. Here, the overall agreement between the calculation and the experiment is  good both in view of orbital character and energetic distance between the peaks, {\em cf.} Fig.\,\ref{fig:xas-dos-pyr}. 

As discussed in Sec.\,\ref{sec:exp} the X-ray absorption spectra turned out to be very different on the oxidized surface, see also Fig.\,\ref{fig:xas-dos-pyr}(b).  Here, the intensity of the Fe $L_3$ edge spectrum is rather independent from the photon incidence angle $\theta$ and the peaks are sharper compared to  Fe OEP on the Cu(001) surface, which coincides with the fact the calculated energy levels are less spread compared to the FeP molecule on the plain Cu(001) surface,  see Fig.\,\ref{fig:xas-dos-pyr}. The $d_{z^2}$ and $d_{\pi}$ orbitals are actually located between the two $d_{x^2-y^2}$ orbitals such that the XAS and the calculated \dos{} are in qualitative agreement. However, in case of the oxidized surface the distance between the two in-plane polarized levels does not match perfectly with the distance between the two main features of the $L_3$ edge which may be related to core-hole effects.

\begin{figure}[tb]
\includegraphics[width=0.485\textwidth]{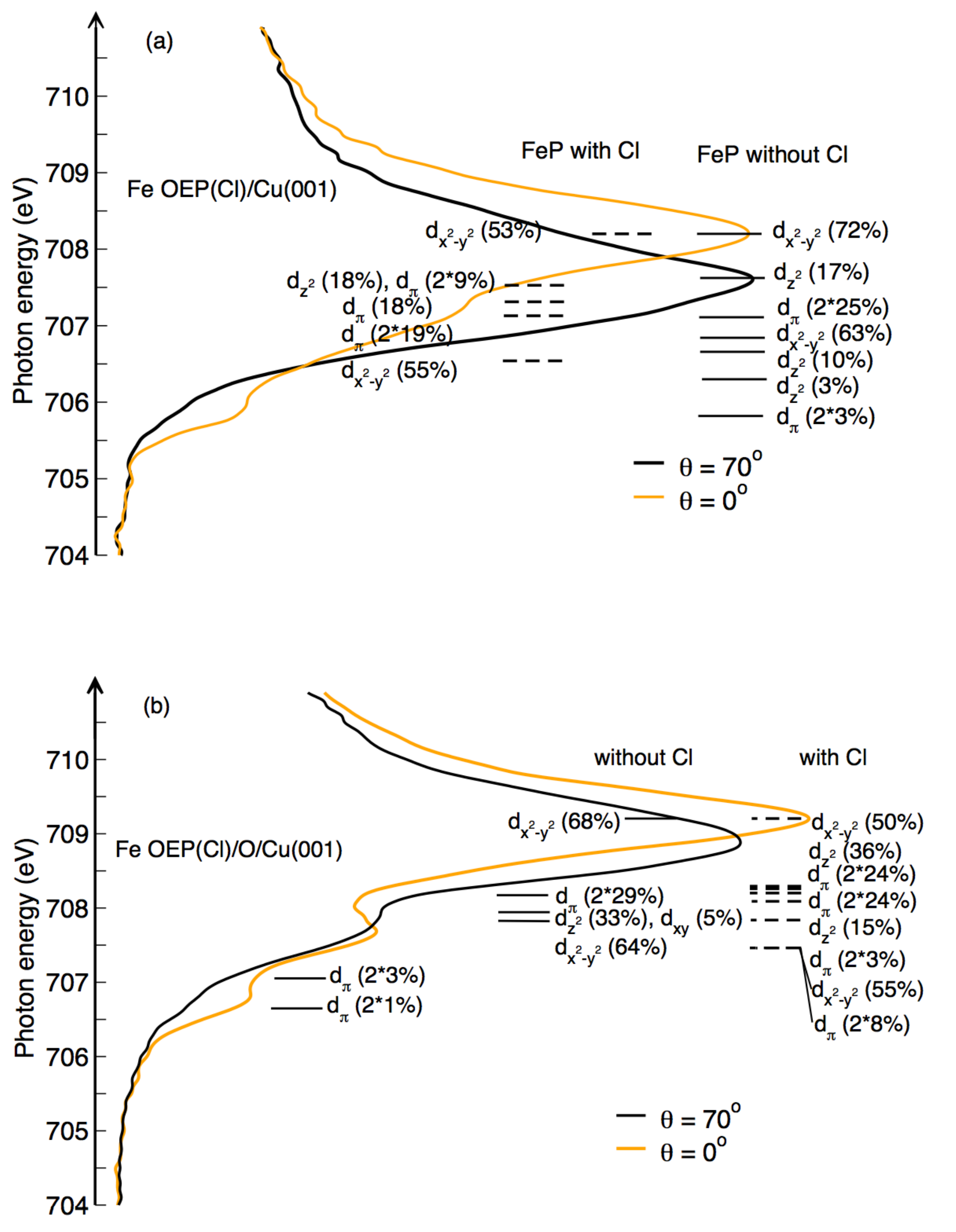}\\
\caption{(Color online) XAS of Fe $L_3$ edge of OEP(Cl) on Cu(001) (a) and   $\sqrt{2}\times2\sqrt{2}\,{\rm R}45^{\circ}$O/Cu(001) (b) measured with linear polarized light. The calculated energy levels of the unoccupied Fe 3$d$ orbitals are shown as horizontal lines. The percentage of the $d$ character is given in brackets. Black lines mark the XAS for normal incidence ($\theta=0^{\circ}$) for which the electric field lies  in the plane and the bright line denotes the XAS obtained for grazing incidence ($\theta = 70^{\circ}$) with the electric field being mostly perpendicular to the surface. } 
\label{fig:xas-dos-Cl}
\end{figure}

The situation becomes more complex if the stabilizing ligand is Cl instead of pyridine, {\em cf.} Fig.\,\ref{fig:xas-dos-Cl}. For Fe OEP(Cl) on Cu(001) the intensity for grazing incidence angles is now larger and lies basically between the features obtained at normal incidence of the photon beam such that the highest intensity is observed for grazing incidence ($\theta = 70^{\circ}$). A comparison of the XAS of the Fe $L_3$ edge of FeP(Cl)/Cu(001) with the energy levels of the unoccupied Fe $d$ states shows that the position of the two $d_{x^2-y^2}$ levels is in reasonable agreement with experiment, whereas the angular dependent intensity of the XAS for grazing incidence barely fits with the calculated $d$ states of FeP/Cu(001), see Fig\,\ref{fig:xas-dos-Cl}. Using the level spectra of FeP(Cl)/Cu(001) instead, the high intensity of the XAS around 708\,eV is quite well reproduced by the broad distribution of $d_{\pi}$ and $d_{z^2}$ states. However, the 2nd $d_{x^2-y^2}$ level is close to 706.5\,eV and therefore, about 0.5\,eV too low. These results can be viewed as an indication that Cl has been only  partially removed and the coexistence of Fe OEP with and without Cl ligand is reflected in the XAS of the Fe $L_3$ edge.

While the XAS of the Fe  $L_{3}$ edge of Fe OEP on Cu(001) shows a significant dependence on the ligand used to stabilize the molecule, this is not observed for the Fe OEP molecule on the oxidized surface, {\em cf.} Fig.\,\ref{fig:xas-dos-pyr} and \ref{fig:xas-dos-Cl}. Independent from the preparation of the Fe OEP molecule the angular dependence of the XAS is quite similar. From a comparison of  the XAS of the Fe $L_3$ edge with the calculated energy levels, it is obvious that the spectrum is already well described by FeP on O/Cu(001). The broadening of the $d_{\pi}$ and $d_{z^2}$ orbitals which occurs due to Cl does not fit with the experimental results. Hence, the influence of Cl seems to be less pronounced in this case. However, more information is needed to underpin the findings from the comparison of XAS and calculated energy spectra. This is done in the next section by analyzing the magnetic properties of the four systems.  

\subsubsection{Effective moments and magnetic anisotropy}\label{sec:aniso}
\begin{figure}[tb]
\includegraphics[width=0.85\columnwidth]{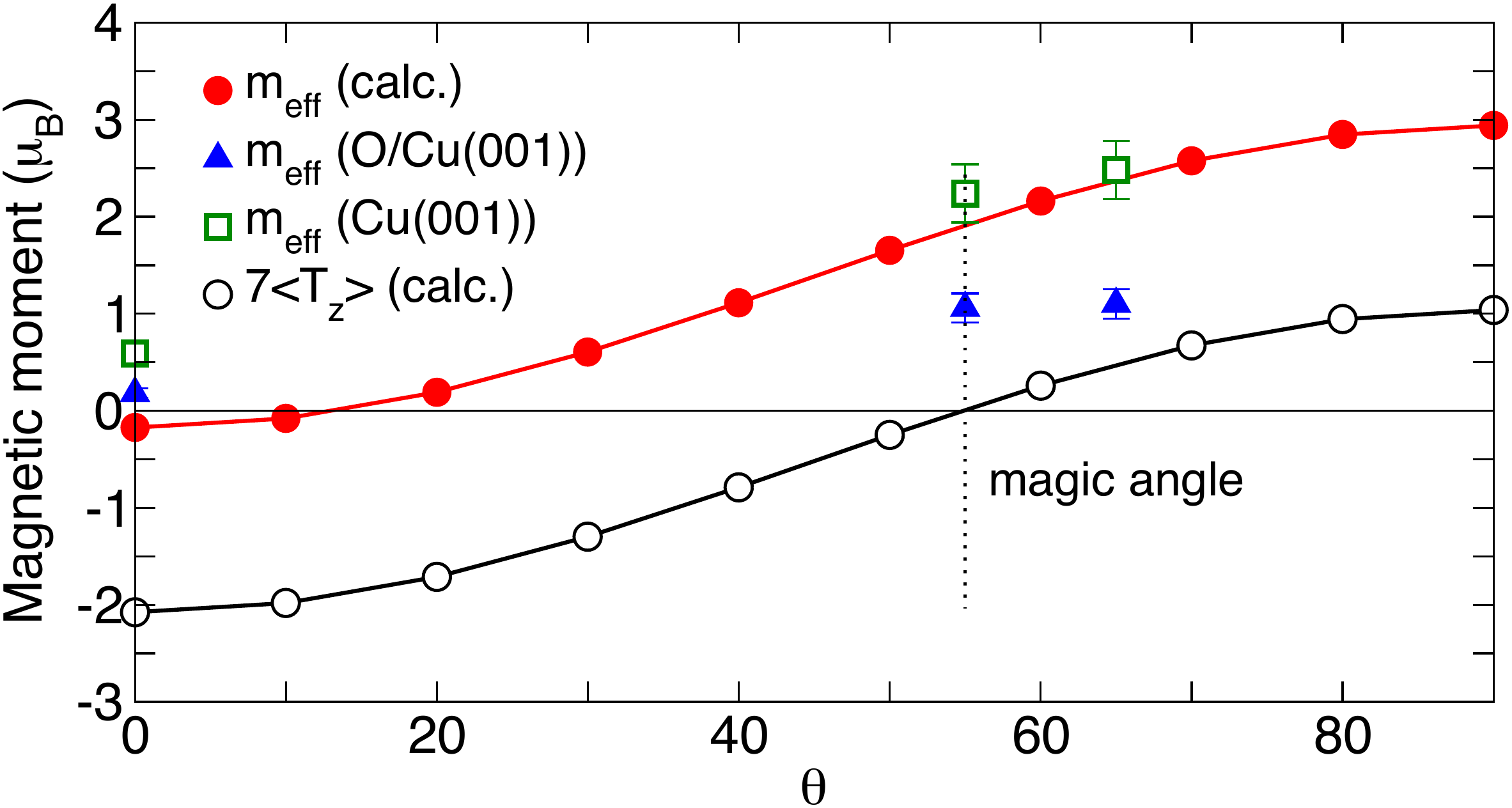}
\caption{(Color online) Effective spin magnetic moments of Fe in FeP/Cu(001) depending on the incidence angle of the photon beam $\theta$. The theoretical effective moments are obtained from the calculated density matrix. The black line denotes the calculated 7$\langle T_z\rangle$ term and the dashed lines marks the magic angle where in saturation  $m_s = m_{\rm eff}$.} 
\label{fig:tz-noO}
\end{figure}

The measured XMCD of the Fe $L_3$ edge of Fe OEP shows a crucial dependence on the photon incidence angle $\theta$, i.e., the signal is very tiny for normal incidence  but large for smaller angles. This is an indication that either the dipole term  $7\langle T_z\rangle$ is large or that the system has a large magneto-crystalline anisotropy. Both quantities are not directly accessible in XAS and XMCD experiments. The spin magnetic moment which is obtained from the sum-rules is an effective spin moment $m_{\rm eff}$ and is related to the actual spin moment $m_s$ by
\begin{equation}
m_{\rm eff}\,=\, m_s \, + \, 7\langle T_z\rangle.
\end{equation}
For bulk systems the dipolar term is usually small and can be neglected, however, it may be significantly large in hybrid systems consisting of molecules and surfaces.\cite{Sipr:09}
To determine $7\langle T_z\rangle$ from the {\em ab initio} calculated density matrix we followed the method of van der Laan.\cite{vanderLaan:98} This model allows the prediction of the effective moment for different photon angles assuming the system is magnetically fully saturated, see Sec.\,\ref{sec:theo}.
The calculated $7\langle T_z\rangle$ term for  FeP on Cu(001) is given in Fig.\,\ref{fig:tz-noO} by the black line and symbols.  
\begin{table}[b]
\caption{Effective spin moments of Fe obtained from sum-rule analysis  and calculated spin moments of Fe $m^{s}_{\rm Fe}$  (including van der Waals forces) in Fe porphyrin on O/Cu(001) and Cu(001) together with the  calculated ratio of spin-up and spin-down charge of the Fe $d$ orbitals and the magnetic anisotropy energy (MAE).  Magnetic moments are given in \mbo{} and energies in meV. The experimentally determined moment are given at the magic angle, whereby not all systems were in saturation, see text.}
\begin{center}
\begin{tabular}{lccccc}\hline\hline
System& $m^{s}_{\rm Fe}$&$m^L_{\rm Fe}/m^s_{\rm Fe}$ &$n_{\uparrow}/n_{\downarrow}$& MAE \\
& (\mbo)&(\%)&& (meV) \\\hline
Experiment:&&&&&\\
Fe\,OEP(Cl)/Cu     & >1.01$\pm$0.15&13&--&--\\
Fe\,OEP(Cl)/O/Cu & >2.32$\pm$0.30&\,\,8&--\\
Fe\,OEP/Cu & >1.06$\pm$0.15&26&&--\\
Fe\,OEP/O/Cu & \,\,2.24$\pm$0.30&12&--&--\\
Theory:&&&&&\\
FeP/Cu &  1.93 & &3.96/2.07&\,0.11\\
FeP/O/Cu & 1.99 &&3.99/2.05&-0.10\\
FeP(Cl)/Cu  &2.50&&4.18/1.69&-0.09\\
FeP(Cl)/O/Cu & 2.62&&4.21/1.64&-0.10\\
FeP(O)/O/Cu & 3.50&&4.66/1.20&\,0.45\\\hline\hline
\end{tabular}
\end{center}
\label{tab:magn}
\end{table}%
Indeed the calculated dipolar term is large, i.e., of the same order as the spin moment itself. In saturation the calculated effective spin moment for grazing  incidence with $\theta = 65^{\circ}$ ($\theta = 90^{\circ}$)  should be 2.45\,\mbo{} (3.0\,\mbo). 
The small negative value obtained for $\theta = 0^{\circ}$ is related to the limited accuracy of the determination of the dipolar term and has no physical meaning. One reason may be related to the fact that the $d_{\pi}$ and $d_{z^2}$ orbitals, which give the main contribution at $\theta = 0^{\circ}$, are less sharp than the in-plane oriented orbitals.
The actual values for $m_{\rm eff}$ obtained from X-ray absorption spectroscopy are much smaller namely 1.06$\,\mu_{\rm B}$ at the magic angle (see Tab.\,\ref{tab:magn}), because the maximum magnetic field, which was available during the measurement, was 5.9\,T which was not sufficient to saturate the sample, see Fig.\,\ref{fig:mag-FeP-Cu}. The occurrence of large dipolar contributions for systems with reduced symmetry has already been predicted by Crocombette {\em et al.}\,\cite{Crocombette:96} who investigated the dipolar term for $3d$ transition metals with different local symmetry.  For  FeP on Cu(001) it turned out that the  calculated $T_z$ values, which reach from -2 to +1\,\mbo, are of the same order as the spin moment $m_s = 1.95$\,\mbo. This agrees qualitatively with recent findings by Stepanow {\em et al.} for Cu phthalocyanine on Ag(001).\cite{Stepanow:10} 

In the presence of  oxygen on the Cu(001) surface the dipolar term, or more precisely,  its angular dependence is almost identical (not shown here). The magnetic spin moment for $\theta = 90^{\circ}$ amounts to 3\,$\mu_{\rm B}$ as in the case without O. This behavior is expected because the occupation of the Fe $d$ levels is very similar in both cases, see Sec.\,\ref{sec:mag}, which is reflected in the density matrices. 
In this case the experimental results for  Fe OEP (Py) on the oxidized surface agree well with the theoretically obtained effective spin moments, see Fig.\,\ref{fig:tz-noO}, because the sample is almost saturated for the maximal applied field of 5.9\,T. At the magic angle both experiment and theory give a spin magnetic moment of 2\,\mbo. The good agreement between experimental and calculated effective spin moments at the magic angle suggests the assumption that the angle dependent changes in the XMCD are not related to a large MAE.

\begin{figure}[tb]
\includegraphics[width=0.85\columnwidth]{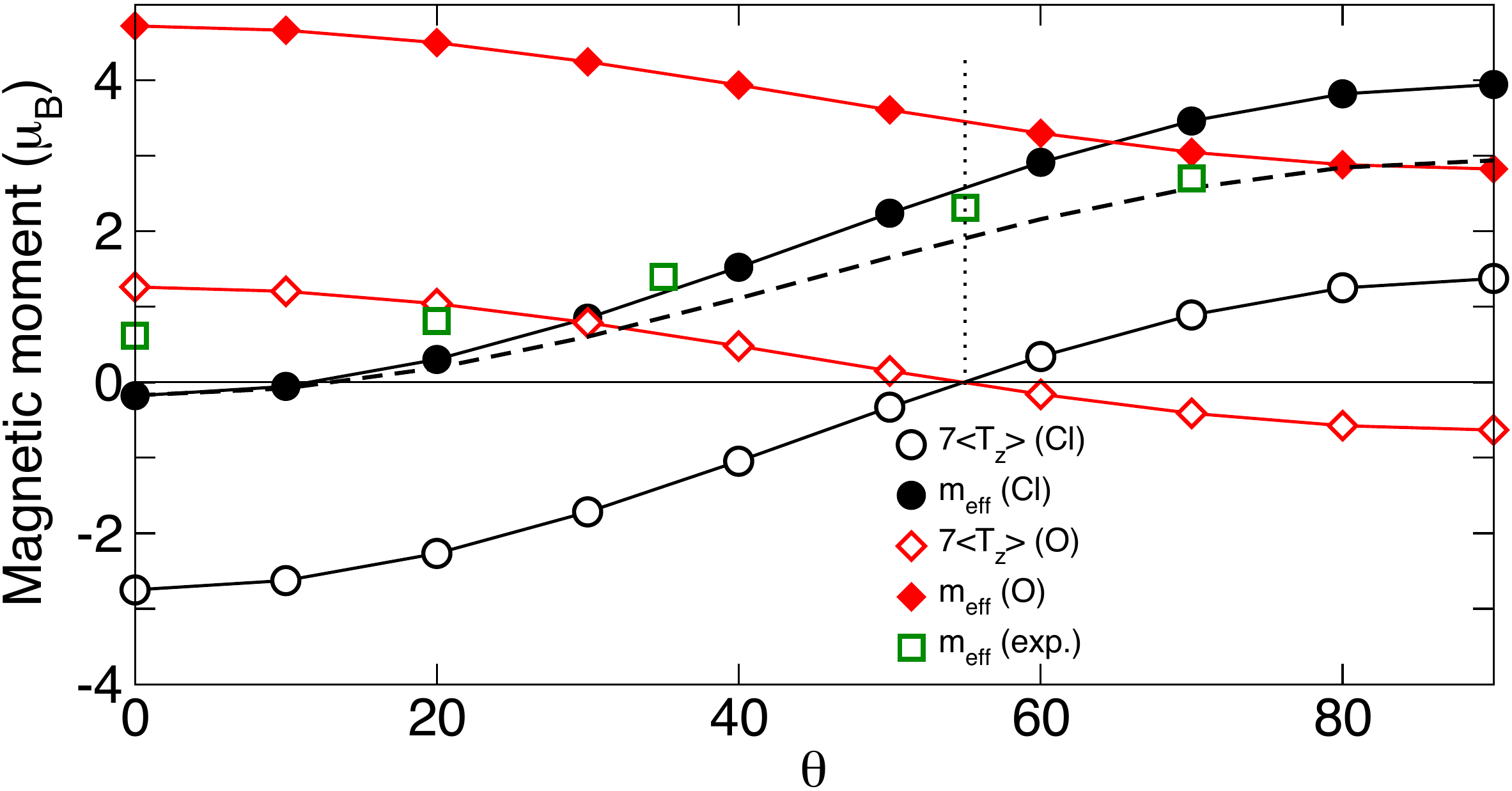}
\caption{(Color online) Calculated effective magnetic spin moments of Fe and dipolar term 7$\langle T_z\rangle$ depending on the incidence angle of the photon beam $\theta$ in the presence of different ligands at the Fe-atom in FeP/$\sqrt{2}\times2\sqrt{2}\,{\rm R}45^{\circ}$/O/Cu(001). Diamonds denote an O-ligand and circles correspond to a Cl ligand. The dashed curve corresponds to the calculated $m_{\rm eff}$ without ligands. The green squares mark the experimental data for Fe OEP(Cl)O/Cu(001).}
\label{fig:tz-withO}
\end{figure}

The dependence on the photon incidence angle is quite different for the Fe OEP  molecules which have been stabilized with Cl ligands. The effective moment obtained from experiment at the magic angle is 2.70\,\mbo{} whereby corresponding to Fig.\,\ref{fig:mag-FeP-Cu} the sample is not fully saturated, see squares in Fig.\,\ref{fig:tz-withO}, and the calculated $m_{\rm eff}$ without ligand clearly underestimates the magnetic moment, {\em cf.} dashed line in Fig.\,\ref{fig:tz-withO}. However, the calculations for  FeP with Cl result in an  effective spin moment which is larger than the experimental value whereas without Cl the effective moment is below the experiment. This underpins our previous observations that Cl has not been completely removed, {\em cf.} Sec.\,\ref{sec:exp-details}. The remaining Cl  has  obviously a significant influence on the magnetic configuration. Oxygen ligands lead to a completely different angular dependence and can be ruled out as possible cause for the different magnetic structures of Py- and Cl-ligated Fe OEP molecules on Cu(001) and O/Cu(001).

However, a huge dipolar term is only one possible reason for the strong angular dependence of the XMCD (Fig.\,\ref{fig:XAS}), the other possible explanation would be a large magnetic anisotropy of the system. Here, we have used the method proposed by Wang\,\cite{Wang:93} to obtain the anisotropy energy from 2nd order perturbation theory. The calculated magnetic anisotropy energies obtained for the porphyrin molecule on the plain and the oxidized  Cu(001) surface are -0.096 (0.114)\,eV with (without) O (minus sign means out-of-plane easy axis). 
The  MAEs are by a factor of 10 smaller than the values obtained by Wang {\em et al.} for Fe phthalocyanine molecules, which have a structure very similar to the FeP molecules investigated here. Wang {\em et al.} calculated for the gas phase molecule an in-plane anisotropy of 1.18\,meV.\cite{Wang:09} However, these calculations were performed within the GGA and therefore, the distance between occupied and unoccupied states was underestimated, which in turn leads to larger anisotropy values, see Sec.\,\ref{sec:theo}.
The MAE obtained for the FeP molecules on Cu(001) is also smaller than for free Jahn-Teller-distorted Fe$_{13}$ clusters, in which the calculated magneto-crystalline anisotropy is about 0.6\,meV/Fe atom.\cite{Sahoo:10} Also in Fe-Pt nanoparticles which have been capped with Cu the magneto-crystalline anisotropy amounts only to 0.5\,meV/atom.\cite{Antoniak:11}  Compared to that the calculated MAE for  FeP on the Cu(001) surface are quite small and the strong changes in the XMCD spectra cannot be explained from the magneto-crystalline anisotropy, but may be related to the dipolar term and its angular dependence.  Unfortunately, due to the large super cells the accuracy of the calculations is limited such that  the determination of MAEs is already at the edge of the accuracy. This means that for small anisotropies the determination of the easy axis is difficult. Except for FeP on Cu(001) the calculations indicate an out-of-plane easy axis, which is not supported by the experimental findings, see Fig.\,\ref{fig:mag-FeP-Cu}. 
It should be mentioned that the influence of the van der Waals forces on the magnetic anisotropy energy is negligible in our case, since for the present system the magnetic configuration remains nearly unchanged if van der Waals interaction is taken into account. Analogous to the previous investigations without ligands, we calculated the effective spin moment and the dipolar term $\langle T_z\rangle$. In the presence of Cl ligands, the dipolar term would be even more important than before. Our calculations of $T_z$ reveal that $m_{\rm eff}$ strongly depends on $\theta$, i.e., the dipolar term is huge being -2.6\,$\mu_{\rm B}$ for normal incidence and 1.4\,$\mu_{\rm B}$ at $90^{\circ}$, {\em cf.} Fig.\,\ref{fig:tz-withO}. The magnetic anisotropy obtained from 2nd order perturbation theory for the Cl-ligated FeP on the O/Cu(001) (Cu(001)) surface  is -0.10\,eV (-0.09\,eV), i.e., it is of the same order of magnitude as for the systems without Cl.  

Although the O ligand case is somewhat artificial, we have also determined the dipolar term and the effective magnetic moment.  The angular dependence of $T_z$ ranges now from -0.63 to 1.25\,$\mu_{\rm B}$, which is small compared to the previous cases without ligand and Cl-ligated FeP molecules, see Fig.\,\ref{fig:tz-withO}. However, the most striking difference is the sign of the $T_z$ term. The average dipolar term is positive being 1.26\,$\mu_{\rm B}$ and has the same sign as in the magnetic spin-moment. The difference arises from different occupation of the spin-down orbitals. Here,  mainly $d_{z^2}$ and $d_{\pi}$ orbitals are occupied, whereas in case of Cl  the minority channel has mainly  $d_{xy}$ character. In this case the magnetic anisotropy from 2nd order perturbation theory amounts to 0.45\,meV which is by a factor of 4 larger than for all other cases. 
\section{Conclusion} \label{sec:concl}
We have presented a combined experimental and theoretical study of the magnetic and electronic properties of Fe OEP molecules on  nonmagnetic substrates. 
Fe OEP molecules on metallic Cu(001) and an oxidized Cu(001) surface have been investigated in view of  the spin-state, the MAE, and the influence of ligands  on the magnetic and electronic properties of the molecule-surface hybrid system.  XAS and XMCD have been measured for different incidence angles of the photon beam. In atmosphere Fe OEP molecules are stabilized by ligands which bond to the Fe atom. Here we have used two different samples one with atomic Cl and one with pyridine, which turned out to have a significant impact on the XAS.

The  spin state and the adsorption position of the molecules on fcc Cu(001) and on the  $\sqrt{2}\times2\sqrt{2}\,{\rm R}45^{\circ}$O/Cu(001) surface have been obtained from DFT calculations. To account for  correlation effects, van der Waals forces were included in the approximation of Grimme. Here, the inclusion of the long-range  interactions basically reduces the molecule-surface distance by 0.5\,\AA. In contrast to adsorption on magnetic surfaces\,\cite{Bernien:09} the molecules are weakly bonded, i.e., the hybridization with the surface is small, however, \dos{} as well as XAS show distinct differences, which are related to the structure of the substrate. Also the interaction mechanism with the surface changes. In case of Cu(001) the hybridization effects and  transfer effects indicate that interaction between Fe atom and surface is indirect via the N atoms, which is usually observed on magnetic metal surfaces. In the presence of the oxygen layer a significant part of the hybridization and interaction with the surface is mediated by the oxygen atoms.
For a quantitative comparison of the effective spin moments obtained from sum-rule analysis of XAS and calculated  spin moments, we determined the dipole term which connects the two quantities.

From DFT calculations an intermediate spin state  with 2\,\mbo{} is obtained for FeP molecules on both substrates, whereas the absorption sites are different, namely missing-row position with oxygen and hollow site position on the unreconstructed metal surface.  The magnetic moment and the spin state are in agreement with the experimental results with pyridine as stabilizing ligand in air, whereby only in the case of the oxidized surface  magnetic saturation could be approached.  The calculated magneto-crystalline anisotropy energy is about 0.1\,meV such that the strong angular dependence of the XMCD has to be mainly related to a huge dipolar term, which is supported by our calculations.

If Cl has been used to stabilize the Fe OEP molecule,  XAS and XMCD show a different dependence on the photon incidence angle, whereby the effect is more expressed on the Cu(001) surface. A comparison of the Fe $L_3$ edge XAS with the energy levels of the unoccupied Fe 3$d$ orbitals gave rise to the assumption that Cl has not been completely removed. This was underpinned by the analysis of the $T_z$ term and the calculation of the effective spin moments.

Summarizing, the combination of X-ray absorption spectroscopy with DFT calculations and, in particular, the analysis of the $T_z$ term  give an accurate picture of the electronic and magnetic structure of Fe porphyrin molecules on  the nonmagnetic substrate and help to distinguish effects from the surface and remaining ligands attached to the molecule.

\section{Acknowledgments}
This work has been supported by the DFG in the context of the Emmy-Noether program (CZ 183/1-1) as well as the SFB 658 and the SFB 491. O.E. acknowledges support from the ERC (project  - ASD), the KAW foundation and the Swedish Research Council (VR).
The Swedish National Infrastructure for Computing (SNIC) is acknowledged to allocate time in high performance supercomputers. Ms. Rita Friese is kindly acknowledged for DSC measurements. 

\begin{thebibliography}{10}
\bibitem{note-Sams}
This observation is in line with the report by Sams and coworkers
  (Ref.\onlinecite{Dolphin:76}) who isolated free Fe(OEP) by heating
  Fe(OEP)Py$_2$ in vacuo at 150\,$^{\circ}$C for 2.5 h.
\end{thebibliography}

\end{document}